\documentclass[preprint,12pt,a4paper,sort&compress,1p,number]{elsarticle}
\usepackage{dcolumn}
\usepackage{graphicx}
\usepackage{caption}
\usepackage{url}
\usepackage{color,hyperref}
\usepackage[T1]{fontenc}
\usepackage{ae,aecompl}
\usepackage{pdflscape}
\usepackage{longtable}
\usepackage{amssymb}
\usepackage{lineno}
\usepackage{rotating}

\usepackage{epsf}
\usepackage{amsmath}
\usepackage{lscape}	
\usepackage{bm}

\usepackage{dcolumn}
\newcolumntype{d}[1]{D{.}{.}{#1}}

\newcommand{\cm}{cm$^{-1}$}
\newcommand{\icm}{cm\textsuperscript{-1}}
\newcommand{\ai}{\textit{ab initio}}
\newcommand{\xa}{$\tilde{X}^{1}A_{1}$}
\newcommand{\ab}{$\tilde{A}^{1}B_{1}$}
\newcommand{\ba}{$\tilde{B}^{1}A_{1}$}

\newcommand{\sihh}{SiH$_{2}$}

\newcommand{\lml}{CATS}

\newcommand{\vecb}[1]{{\mathbf #1}}

\newcommand{\p}{^\prime}

\newcommand{\TROVE}{{\sc TROVE}}

\newcommand{\Cv}[1]{${\mathcal C}_{#1{\rm v}}$}

\journal{JQSRT}
\begin{document}
\begin{frontmatter}

\title{The high-temperature rotation-vibration spectrum and rotational clustering of silylene (SiH$_2$)}

\author{Victoria H. J. Clark}
\author{Alec Owens}
\author{Jonathan Tennyson}
\author{Sergei N. Yurchenko}
\ead{s.yurchenko@ucl.ac.uk}
\address{Department of Physics and Astronomy, University College London, Gower Street, WC1E 6BT London, United Kingdom}

\begin{abstract}
A rotation-vibration line list for the electronic ground state (\xa) of \sihh\ is presented.  The line list, named \lml, is suitable for temperatures up to 2000~K and covers the wavenumber range 0--10\,000~\cm\ (wavelengths $>1.0$~$\mu$m) for states with rotational excitation up to $J=52$.  Over 310 million transitions between 593\,804 energy levels have been computed variationally with a new empirically refined potential energy surface, determined by refining to 75 empirical term values with  $J\leq 5$ and a newly computed high-level \ai\ dipole moment surface. This is the first, comprehensive high-temperature line list to be reported for SiH$_2$ and it is expected to aid the study of silylene in plasma physics, industrial processes and possible astronomical detection. Furthermore, we investigate the phenomenon of rotational energy level clustering in the spectrum of SiH$_2$. The \lml\ line list is available from the ExoMol database (\href{www.exomol.com}{www.exomol.com}) and the CDS database.
\end{abstract}


\end{frontmatter}

\section{Introduction}

The spectrum of \sihh, known as silylene or silicon dihydride, was first observed in the late 1960s~\citep{67DuHeVe.SiH2} and, since then, numerous experimental and theoretical studies have followed (see Ref.~\citep{16KoMaSt.SiH2} and references within). Its importance in silicon chemistry is largely owing to it being an intermediate in many chemical reactions involving silane (SiH$_4$). For example, the formation of \sihh\ is used to monitor the decomposition of SiH$_4$ into silylene and hydrogen; and the loss of \sihh\ can be used to track the formation of disilane (Si$_2$H$_6$)~\citep{87JaMeSc.SiH2, 90DoDoGa.SiH2, 88JaChxx.SiH2}. Given the widespread use of silane plasmas in industry, detailed spectroscopic information on SiH$_2$ can significantly help the detection and monitoring of certain processes, such as the deposition of hydrogenated amorphous silicon (a-Si:H) films, the chemistry of which is not fully understood~\citep{19NaPiDa.SiH3}. Silylene is yet to be conclusively detected astronomically with unsuccessful searches in the circumstellar envelope of the carbon star IRC+10216~\citep{92Turner.SiN,94AvBeCu.SiH2}, despite speculation of its presence in these environments~\citep{99McChxx.SiH2}.

Theoretical work on SiH$_2$ has primarily focused on \textit{ab initio} predictions of the electronic state ordering relative to the isovalent methylene (CH$_2$), which interestingly has a triplet ground state (see Refs.~\citep{03ApPaKa.SiH2,04KaDuMa.SiH2} and references within). More closely related to this study are the full-dimensional variational calculations of the ro-vibrational and ro-vibronic spectra of the ground and excited singlet states, \xa, \ab\ and \ba~\citep{93GaRoYa.SiH2,04YuBuKr.SiH2,05GuBuJe.SiH2}. Nowadays, variational approaches are more robust and can offer high-accuracy predictions of line positions and intensities over extended wavenumber ranges, capable of supporting high-resolution spectroscopic measurements~\citep{jt654}. In particular, they have found widespread application in exoplanetary science through the ExoMol database~\citep{jt541,jt631}, which provides molecular line lists on a large variety of small molecules relevant to the atmospheric characterization of exoplanets and other hot bodies. In
particular the ExoMol database already contains line lists for a number of silicon-bearing molecules: SiH$_4$~\citep{jt701}, SiH~\citep{jt776}, SiO~\citep{jt563} and SiS~\citep{jt724}.
It is within the ExoMol computational framework~\citep{jt626} that we treat silylene.

In this paper, we present a comprehensive molecular line list for the ground electronic state \xa\ of SiH$_2$. The line list, named \lml, has been computed using robust first-principles methodologies with a degree of empirical tuning to the available spectroscopic data, namely the refinement of a newly computed high-level \textit{ab initio} potential energy surface (PES) to experimentally determined ro-vibrational term values. The CATS line list is applicable for temperatures up to $T=2000$~K and contains over 254 million transitions between states with rotational excitation up to $J=51$.

Interestingly, triatomic symmetrical hydrides of the form XH$_2$ with a heavy central nuclei X and an interbond angle close to 90$^{\circ}$ exhibit well separated, near-degenerate rotational energy level clusters at high rotational excitation~\citep{00Jensen.cluster}. This effect arises in local-mode molecules when the bonds are nearly orthogonal to each other such that the rotation of the molecule is not destabilized by Coriolis-type interactions with the vibrational modes. The new CATS line list is used to explore this phenomenon in SiH$_2$, which has an interbond angle $\alpha_{\rm e}=92.04^{\circ}$~\citep{16KoMaSt.SiH2} and is expected to form rotational cluster states at high rotational excitation.

The paper is structured as follows: In Sec.~\ref{sec:theory} we describe the theoretical approach including the construction of the PES and subsequent empirical refinement, the dipole moment surface (DMS) and the variational calculations used to produce the line list. In Sec.~\ref{sec:results}, the line list is presented and evaluated along with analysis of the temperature-dependent partition function. In Sec.~\ref{s:cluster}, rotational energy level clustering in SiH$_2$ is investigated. Conclusions are offered in Sec.~\ref{sec:conc}.

\section{Theoretical approach}
\label{sec:theory}

\subsection{Potential Energy Surface}

The initial \ai\ PES was computed using the explicitly correlated coupled cluster method CCSD(T)-F12c~\citep{10HaTeKo.ai} with the F12-optimized correlation consistent basis set, cc-pVQZ-F12~\citep{08PeAdWe.ai} in the frozen core approximation. Calculations employed the diagonal fixed amplitude ansatz 3C(FIX)~\citep{04TenNo.ai} and a Slater geminal exponent value of $\beta=1.0$~$a_0^{-1}$~\citep{09HiPeKn.ai}. The auxiliary basis sets were chosen to be the resolution of the identity OptRI~\citep{08YoPexx.ai} basis and the cc-pV5Z/JKFIT~\citep{02Weigend.ai} and aug-cc-pwCV5Z/MP2FIT~\citep{05Hattig.ai} basis sets for density fitting. MOLPRO2015~\citep{MOLPRO} was used for all electronic structure calculations. The PES was computed on a uniformly-spaced grid of 1898 nuclear geometries with energies up to $h c \cdot 19\,300$~\cm\ ($h$ is the Planck constant and $c$ is the speed of light). The grid was built in terms of three internal coordinates: the two Si--H bond lengths $1.2\leq r_{\rm SiH_1},r_{\rm SiH_2} \leq 2.0$~\AA\ and the interbond angle $30\leq \alpha({\rm H_{1}SiH_{2}})\leq 160^{\circ}$.

The three coordinates of the PES were chosen to be,
\begin{eqnarray}
y_1 &=& 1-\exp\left(-b (r_1-r_{\rm e})\right), \\
y_2 &=& 1-\exp\left(-b (r_2-r_{\rm e})\right), \\
y_3 &=& \cos\alpha-\cos\alpha_{\rm e},
\end{eqnarray}
where $r_1$ and $r_2$ are the bond lengths Si-H$_1$ and Si-H$_2$, respectively,  $b$ is the Morse parameter, $\alpha $ is the interbond angle $\angle $(H-Si-H), and $r_{\rm e}$ and $\alpha_{\rm e}$ are the corresponding equilibrium parameters. The PES was represented using the analytical function~\citep{01TyTaSc.H2S},
\begin{equation}\label{e:PES}
  V =   V_{0} +  V_{\rm HH},
\end{equation}
where
\begin{equation}
\label{e:V0}
V_0 =  \sum_{0 \le i+j+k\le 6} f_{ijk} y_1^i y_2^j y_3^k ,
\end{equation}
and
\begin{equation}
\label{e:Vhh}
V_{\rm HH} = B_1 \exp(-g_1 r_{\rm HH}) + B_2 \exp(-g_2 r_{\rm HH}^2) .
\end{equation}
The $f_{ijk}$ expansion parameters obey the symmetric relation $f_{ijk} = f_{jik}$ owing to the identical properties of the hydrogen atoms 1 and 2. The distance between the hydrogen nuclei
\begin{equation}
r_{\rm HH}=\sqrt{r_1^2+r_2^2-2r_1r_2\cos\alpha},
\end{equation}
and the numerical parameters $B_1$, $B_2$, $g_1$ and $g_2$ are taken as in Ref.~\citep{01TyTaSc.H2S} (see supplementary material for values). The contribution $V_{\rm HH}$ is to prevent holes appearing in the PES at geometries where $r_{\rm HH}$ is small.
 The \ai\ data was weighted with factors of the form~\citep{97PaScxx.H2O}
\begin{equation}\label{eq:weights}
w_i=\left(\frac{\tanh\left[-0.0008\times(\tilde{E}_i - 12{\,}000)\right]+1.002002002}{2.002002002}\right) .
\end{equation}
Watson's robust fitting scheme~\citep{03Watson.methods} was also utilized to reduce the weights of outliers and improve the description at lower energies. A total of 53 parameters were varied (46 expansion parameters + 2 equilibrium parameters + 1 Morse parameter + 4 damping parameters) in a least-squares fitting to the \ai\ data which were reproduced with a weighted root-mean-square (rms) error of 0.0048~cm$^{-1}$.

\subsection{Dipole Moment Surface}

To represent the instantaneous dipole moment vector $\bm{\mu}$ of SiH$_2$, we employed the so-called $pq$ axis system~\citep{88Jensen.CH2}. The $p$ and $q$ axes were defined in the plane of the three nuclei with origin at the Si atom. The $q$ axis bisects the interbond angle and the $p$ axis lies perpendicular to the $q$ axis. In electronic structure calculations, an external electric field with components $\pm0.005$~a.u. was applied along each axis and the respective dipole moment component $\mu_{p}$ and $\mu_q$ determined using central finite differences. Calculations were at the same level of theory as the PES, namely CCSD(T)-F12c/cc-pVQZ-F12 and used the same grid of 1898 nuclear geometries.

In order to represent the two dipole components analytically the following expansions were used \citep{93JoJexx.H2O} (see also Ref.~\citep{jt607}),
\begin{eqnarray}
 \mu^{(q)} &=&  \sin\alpha \sum_{i+j+k\le 5} F^{(q)}_{ijk} \xi_1^i \xi_2^j \xi_3^k, \\
  \label{e:muQ:MORBID}
 \mu^{(p)} &=&  \sum_{i+j+k\le 5} F^{(p)}_{ijk} \xi_1^i \xi_2^j \xi_3^k,
  \label{e:muP:MORBID}
\end{eqnarray}
where the coordinates
\begin{eqnarray}
  \xi_1 &=& r_1-r_{\rm e}, \\
  \xi_2 &=& r_2-r_{\rm e},\\
  \xi_3 &=& \cos\alpha-\cos\alpha_{\rm e}.
\end{eqnarray}
The dipole expansion parameters $F^{(q)}_{ijk}$ and $F^{(p)}_{ijk}$ are subject to the conditions that the function $\mu^{(q)}$ is unchanged under the interchange of the identical protons, whereas the function $\mu^{(p)}$ is antisymmetric under this operation. The same analytic representation of the DMS was first used for CH$_{2}$ \citep{88Jensen.CH2} and also previously for \sihh\ \citep{04YuBuKr.SiH2}.

The dipole expansion parameters were determined through a least-squares fitting to the \ai\ data. The same weight factors as given in Eq.~\eqref{eq:weights} were used along with Watson's robust fitting scheme and the equilibrium parameters fixed to $r_{\rm e} = 1.514$~\AA\ and $\alpha_{\rm e} =92.0^\circ$. The $\mu_p$ component required 20 parameters (including the 2 equilibrium parameters) and reproduced the \textit{ab initio} data with a weighted rms error of   $1.5\times 10^{-4}$~D with the $\mu_q$ component using 26 parameters (including the 2 equilibrium parameters) giving a weighted rms error of $1\times 10^{-3}$~D. The DMSs of SiH$_2$ are provided in the supplementary material. Our \ai\ dipole moment at the equilibrium (0.066~D) is in a good agreement with the \ai\ equilibrium dipole moment reported by Gabriel \textit{et al.} \citep{93GaRoYa.SiH2} (0.075~D) and slightly smaller than the equilibrium dipole moment from Ref.~\citep{04YuBuKr.SiH2} (0.14 D).

\subsection{Variational calculations}

Variational calculations were performed with the nuclear motion program \TROVE~\citep{TROVE} and details of its methodology are discussed extensively elsewhere~\citep{TROVE,15YaYuxx.method,17YuYaOv.methods,jt626}. Here, we summarise the main calculation steps.

The \TROVE\ kinetic energy operator was constructed as a sixth-order Taylor expansion around the SiH$_2$ equilibrium geometry in terms of the following linearized coordinates,
\begin{eqnarray}
  \chi_1^{\rm lin} &=& r_1^{\rm lin}-r_{\rm e}, \\
  \chi_2^{\rm lin} &=& r_2^{\rm lin}-r_{\rm e}, \\
  \chi_3^{\rm lin} &=& \alpha^{\rm lin}-\alpha_{\rm e} .
\end{eqnarray}
where $r_i^{\rm lin}$ and $\alpha^{\rm lin}$ are the linearized versions of $r_i$ and $\alpha$, respectively (see Ref.~\citep{TROVE}). The PES was also re-expanded to sixth-order in terms of the coordinates
\begin{eqnarray}
  \zeta_1^{\rm lin} &=& 1-\exp(-b (r_1^{\rm lin}-r_{\rm e})), \\
  \zeta_2^{\rm lin} &=& 1-\exp(-b (r_2^{\rm lin}-r_{\rm e})), \\
  \zeta_3^{\rm lin} &=& \alpha^{\rm lin}-\alpha_{\rm e}.
\end{eqnarray}

For the primitive basis set, \TROVE\ uses 1D numerical basis functions $\phi_{n_1}(\chi_1)$, $\phi_{n_2}(\chi_2)$ and $\phi_{n_3}(\chi_3)$ constructed with the Numerov-Cooley approach~\citep{24Numerov.method,61Cooley.method}. The eigenfunctions of the 1D stretching and bending Hamiltonian operators $\hat{H}_{i}^{\rm (1D)}$ were obtained by freezing all other degrees of freedom at their equilibrium values. In order to improve the primitive basis set by making it more compact, a two-step contraction scheme is used. At step~1, the 1D basis functions are combined into two subgroups, one for the stretches
 \begin{equation}\label{e:class1}
 \phi^{\rm (2D)}_{n_1,n_2}(\chi_1, \chi_2) = \phi_{n_1}(\chi_1) \phi_{n_2}(\chi_2),
 \end{equation}
 and the other for the bending mode,
 \begin{equation}\label{e:class2}
 \phi^{\rm (1D)}_{n_3}(\chi_1) = \phi_{n_3}(\chi_3)
 \end{equation}
and these are used to solve the respective reduced Hamiltonian operators (stretching $\hat{H}^{\rm (2D)}$ and bending $\hat{H}^{\rm (1D)}$). The reduced Hamiltonians are constructed by averaging the total vibrational Hamiltonian operator $\hat{H}^{(J=0)}$ over the other ground vibrational basis functions~\citep{18ChJeYu.C2H2,jt730}. The resulting eigenfunctions of the two reduced problems, $\psi^{\rm (2D)}_{\lambda_1,\lambda_2}$ and $\psi^{\rm (1D)}_{\lambda_3}$, are contracted and classified according to the \Cv{2}(M) symmetry group~\citep{98BuJexx} using an optimized symmetrization procedure~\citep{17YuYaOv.methods} to form a symmetry-adapted 3D vibrational basis set as a product  $\psi^{\rm (2D)}_{\lambda_1\lambda_2} \psi^{\rm (1D)}_{\lambda_3}$. This vibrational basis set is then used for the $J=0$ eigenproblem at step 2, with the eigenfunctions contracted again and used to form the symmetry-adapted $J>0$ ro-vibrational basis set.

In steps 1 and 2 an additional basis set cut-off was applied based on the polyad number,
\begin{equation}
P = 2 (n_1 + n_2) + n_3 \leq P_{\rm max}
\end{equation}
which used the polyad cutoff $P_{\rm max} = 24$. The maximal values of $n_1, n_2$ and $n_3$ that define the size of the primitive basis set were 12, 12 and 24, respectively. The contracted $J=0$ basis set contained 192 and 143 basis functions with energies up to 16\,000~\cm\ for the $A_1$ and $B_2$ symmetries, respectively. For the rotational basis set, symmetrized spherical harmonics were used~\citep{17YuYaOv.methods}.

Line list calculations employed the empirically refined PES (discussed below) and \ai\ DMS. All transitions and corresponding line strengths were computed for the 0--10\,000~\cm\ range with a lower state energy threshold of 10\,000~\cm\ and upper state threshold of 18\,000~\cm. States were computed up to $J=60$ but only those below $J=51$ contributed to the line list with the $J>51$ energies lying above 10\,000~\cm. The nuclear spin statistical weights of $^{28}$Si$^{1}$H$_2$ are $g_{(\rm ns)}=\lbrace 1,1,3,3\rbrace$ for states of symmetry $\lbrace A_1,A_2,B_1,B_2\rbrace$, respectively. Atomic mass values were used for the \TROVE\ calculations.

\subsection{Potential energy surface refinement}

To improve the accuracy of the computed line list the initial \ai\ PES was empirically refined using an efficient least-squares fitting procedure~\citep{jt503} implemented in \TROVE. Experimental term values up to $J=5$ were extracted from the literature~\citep{89YaKaHi.SiH2,91IsKaxx.SiH2,99HiIsxx.SiH2,02IsMuMi.SiH2} and these are listed in Table~\ref{t:obs-calc}. Since spectroscopic data for the ground electronic state is very limited, additional term values of the $\nu_{2}$ bending mode were generated using the spectroscopic constants from Ref.~\citep{89YaKaHi.SiH2} with the PGOPHER program~\citep{PGOPHER}. Pure rotational energies up to $J=5$ were generated in a similar manner using the spectroscopic constants of Ref.~\citep{89YaKaHi.SiH2}. We mention that in some instances we have extracted the `perturbed' band origins, i.e. those directly observed in experiment and not the `unperturbed' values fitted in the analysis along with resonance/coupling parameters. For example, in Ref.~\citep{99HiIsxx.SiH2} the perturbed $\nu_1$ fundamental is at 2005.4692~\cm, while the unperturbed band origin is given as 1995.9280~\cm.

\LTcapwidth=\textwidth
{\tiny
\begin{longtable}{@{\extracolsep{0.0cm}} d{6}rrrrrcccccccl}
\caption{\label{t:obs-calc}Results of the potential energy surface (PES) refinement. Observed term values are compared against those computed with the \ai\ and refined PESs. Each energy level is described by seven (standard) quantum numbers  and has been assigned a weight in the refinement (see text). The \TROVE\ assignment of the three entries marked with an $^a$ disagree with the experimental assignment. The term values marked with a $^b$, referenced by \citep{99HiIsxx.SiH2}, are from an unpublished work. Term values from Ref.~\citep{89YaKaHi.SiH2} were generated with their spectroscopic constants using the program PGOPHER~\citep{PGOPHER}. The experimental term values from Ref.~\citep{02IsMuMi.SiH2} were shifted by -3.5~\cm\ (see text).}\\
\hline\hline
\multicolumn{1}{c}{Observed}&  \textit{Ab initio} & Refined & $\Delta_{\rm abinit}$ & $\Delta_{\rm ref}$ & Wt. & $J$ & $\Gamma $ & $K_a$ & $K_c$ & $v_1$ & $v_2$  &$ v_3$ & Ref. \\ \hline
\endfirsthead
\caption{(\textit{Continued})}\\ \hline\hline
\multicolumn{1}{c}{Observed}&  \textit{Ab initio} & Refined & $\Delta_{\rm abinit}$ & $\Delta_{\rm ref}$ & Wt. & $J$ & $\Gamma $ & $K_a$ & $K_c$ & $v_1$ & $v_2$  &$ v_3$ & Ref. \\ \hline
\endhead
     998.6241 &    998.177 &    998.608 &     0.447 &     0.016 &     1.000 &    0 &$ A_1 $&   0 &    0 &   0 &    1 &   0 &  \citep{89YaKaHi.SiH2}    \\
     1992.816 &   1996.464 &   1993.000 &    -3.648 &    -0.184 &     0.100 &    0 &$ B_2 $&   0 &    0 &   0 &    0 &   1 &  \citep{99HiIsxx.SiH2}$^b$\\
    2005.4692 &   2008.831 &   2005.357 &    -3.361 &     0.112 &     0.100 &    0 &$ A_1 $&   0 &    0 &   1 &    0 &   0 &  \citep{99HiIsxx.SiH2}    \\
       2952.7 &   2953.933 &   2951.678 &    -1.233 &     1.022 &     0.010 &    0 &$ A_1 $&   0 &    0 &   0 &    3 &   0 &  \citep{91IsKaxx.SiH2}$^{a}$\\
       2998.6 &   3001.685 &   2998.785 &    -3.085 &    -0.185 &     0.010 &    0 &$ A_1 $&   0 &    0 &   1 &    1 &   0 &  \citep{91IsKaxx.SiH2}$^{a}$\\
       3907.4 &   3914.162 &   3907.789 &    -6.762 &    -0.389 &     0.010 &    0 &$ A_1 $&   0 &    0 &   1 &    2 &   0 &  \citep{91IsKaxx.SiH2}    \\
       3923.3 &   3930.305 &   3923.642 &    -7.005 &    -0.342 &     0.010 &    0 &$ A_1 $&   0 &    0 &   2 &    0 &   0 &  \citep{91IsKaxx.SiH2}    \\
       3976.8 &   3981.070 &   3975.988 &    -4.270 &     0.812 &     0.010 &    0 &$ B_2 $&   0 &    0 &   1 &    0 &   1 &  \citep{91IsKaxx.SiH2}$^{a}$\\
       3997.5 &   4004.096 &   3996.334 &    -6.596 &     1.166 &     0.010 &    0 &$ A_1 $&   0 &    0 &   0 &    0 &   2 &  \citep{91IsKaxx.SiH2}    \\
      1010.64 &   1010.205 &   1010.627 &     0.433 &     0.011 &     1.000 &    1 &$ A_2 $&   1 &    1 &   0 &    1 &   0 &  \citep{02IsMuMi.SiH2}    \\
      1990.12 &   1992.097 &   1990.346 &    -1.978 &    -0.228 &     0.010 &    1 &$ A_2 $&   1 &    1 &   0 &    2 &   0 &  \citep{02IsMuMi.SiH2}    \\
      2017.32 &   2020.688 &   2017.434 &    -3.369 &    -0.115 &     0.010 &    1 &$ A_2 $&   1 &    1 &   1 &    0 &   0 &  \citep{02IsMuMi.SiH2}    \\
      2964.02 &   2966.188 &   2963.896 &    -2.169 &     0.122 &     0.010 &    1 &$ A_2 $&   1 &    1 &   0 &    3 &   0 &  \citep{02IsMuMi.SiH2}    \\
      3010.78 &   3013.822 &   3010.941 &    -3.043 &    -0.162 &     0.010 &    1 &$ A_2 $&   1 &    1 &   1 &    1 &   0 &  \citep{02IsMuMi.SiH2}    \\
      3919.02 &   3926.280 &   3919.667 &    -7.261 &    -0.648 &     0.010 &    1 &$ A_2 $&   1 &    1 &   2 &    0 &   0 &  \citep{02IsMuMi.SiH2}    \\
      3935.95 &   3942.153 &   3935.672 &    -6.204 &     0.277 &     0.002 &    1 &$ A_2 $&   1 &    1 &   0 &    0 &   2 &  \citep{02IsMuMi.SiH2}    \\
      3987.83 &   3993.210 &   3987.986 &    -5.381 &    -0.157 &     0.002 &    1 &$ A_2 $&   1 &    1 &   1 &    0 &   0 &  \citep{02IsMuMi.SiH2}    \\
      4008.23 &   4015.819 &   4008.194 &    -7.590 &     0.035 &     0.002 &    1 &$ A_2 $&   1 &    1 &   0 &    0 &   1 &  \citep{02IsMuMi.SiH2}    \\
      4873.99 &   4880.069 &   4873.381 &    -6.080 &     0.608 &     0.001 &    1 &$ A_2 $&   1 &    1 &   1 &    3 &   0 &  \citep{02IsMuMi.SiH2}    \\
      4901.48 &   4908.368 &   4901.882 &    -6.889 &    -0.403 &     0.001 &    1 &$ A_2 $&   1 &    1 &   2 &    1 &   0 &  \citep{02IsMuMi.SiH2}    \\
      4953.24 &   4959.496 &   4953.837 &    -6.257 &    -0.598 &     0.001 &    1 &$ A_2 $&   1 &    1 &   1 &    1 &   1 &  \citep{02IsMuMi.SiH2}    \\
      4989.73 &   4996.392 &   4990.789 &    -6.663 &    -1.060 &     0.001 &    1 &$ A_2 $&   1 &    1 &   1 &    1 &   1 &  \citep{02IsMuMi.SiH2}    \\
      5819.86 &   5826.200 &   5818.389 &    -6.341 &     1.470 &     0.001 &    1 &$ A_2 $&   1 &    1 &   1 &    4 &   0 &  \citep{02IsMuMi.SiH2}    \\
      5859.73 &   5867.706 &   5860.959 &    -7.977 &    -1.230 &     0.001 &    1 &$ A_2 $&   1 &    1 &   2 &    2 &   0 &  \citep{02IsMuMi.SiH2}    \\
      5967.03 &   5974.871 &   5969.095 &    -7.842 &    -2.067 &     0.001 &    1 &$ A_2 $&   1 &    1 &   0 &    4 &   1 &  \citep{02IsMuMi.SiH2}    \\
      6756.18 &   6764.534 &   6755.532 &    -8.355 &     0.647 &     0.001 &    1 &$ A_2 $&   1 &    1 &   1 &    5 &   0 &  \citep{02IsMuMi.SiH2}    \\
      6808.93 &   6819.310 &   6812.410 &   -10.381 &    -3.481 &     0.001 &    1 &$ A_2 $&   1 &    1 &   2 &    3 &   0 &  \citep{02IsMuMi.SiH2}    \\
      6842.27 &   6854.335 &   6842.308 &   -12.066 &    -0.039 &     0.000 &    1 &$ A_2 $&   1 &    1 &   1 &    3 &   1 &  \citep{02IsMuMi.SiH2}    \\
      6883.65 &   6895.644 &   6883.496 &   -11.995 &     0.153 &     0.000 &    1 &$ A_2 $&   1 &    1 &   2 &    1 &   1 &  \citep{02IsMuMi.SiH2}    \\
      6935.53 &   6944.916 &   6940.619 &    -9.387 &    -5.090 &     0.000 &    1 &$ A_2 $&   1 &    1 &   0 &    5 &   1 &  \citep{02IsMuMi.SiH2}    \\
      11.8007 &     11.809 &     11.793 &    -0.008 &     0.008 &     5.000 &    1 &$ A_2 $&   1 &    1 &   0 &    0 &   0 &  \citep{89YaKaHi.SiH2}    \\
    1010.6389 &   1010.205 &   1010.627 &     0.433 &     0.011 &     1.000 &    1 &$ A_2 $&   1 &    1 &   0 &    1 &   0 &  \citep{89YaKaHi.SiH2}    \\
      10.7248 &     10.666 &     10.721 &     0.059 &     0.004 &     5.000 &    1 &$ B_1 $&   0 &    1 &   0 &    0 &   0 &  \citep{89YaKaHi.SiH2}    \\
    1009.4191 &   1008.914 &   1009.393 &     0.505 &     0.026 &     1.000 &    1 &$ B_1 $&   0 &    1 &   0 &    1 &   0 &  \citep{89YaKaHi.SiH2}    \\
      15.1206 &     15.100 &     15.123 &     0.020 &    -0.003 &     5.000 &    1 &$ B_2 $&   1 &    0 &   0 &    0 &   0 &  \citep{89YaKaHi.SiH2}    \\
    1014.1412 &   1013.679 &   1014.134 &     0.462 &     0.007 &     1.000 &    1 &$ B_2 $&   1 &    0 &   0 &    1 &   0 &  \citep{89YaKaHi.SiH2}    \\
      29.7005 &     29.592 &     29.677 &     0.109 &     0.023 &     5.000 &    2 &$ A_1 $&   0 &    2 &   0 &    0 &   0 &  \citep{89YaKaHi.SiH2}    \\
      45.5642 &     45.530 &     45.569 &     0.034 &    -0.005 &     5.000 &    2 &$ A_1 $&   2 &    0 &   0 &    0 &   0 &  \citep{89YaKaHi.SiH2}    \\
    1028.4507 &   1027.900 &   1028.411 &     0.551 &     0.040 &     1.000 &    2 &$ A_1 $&   0 &    2 &   0 &    1 &   0 &  \citep{89YaKaHi.SiH2}    \\
    1045.4185 &   1044.956 &   1045.436 &     0.462 &    -0.018 &     1.000 &    2 &$ A_1 $&   2 &    0 &   0 &    1 &   0 &  \citep{89YaKaHi.SiH2}    \\
      39.8798 &     39.714 &     39.886 &     0.166 &    -0.006 &     5.000 &    2 &$ A_2 $&   1 &    1 &   0 &    0 &   0 &  \citep{89YaKaHi.SiH2}    \\
    1039.2223 &   1038.615 &   1039.202 &     0.607 &     0.020 &     1.000 &    2 &$ A_2 $&   1 &    1 &   0 &    1 &   0 &  \citep{89YaKaHi.SiH2}    \\
      43.1013 &     43.135 &     43.093 &    -0.033 &     0.008 &     5.000 &    2 &$ B_1 $&   2 &    1 &   0 &    0 &   0 &  \citep{89YaKaHi.SiH2}    \\
    1042.8724 &   1042.479 &   1042.894 &     0.393 &    -0.022 &     1.000 &    2 &$ B_1 $&   2 &    1 &   0 &    1 &   0 &  \citep{89YaKaHi.SiH2}    \\
      29.9299 &     29.849 &     29.904 &     0.081 &     0.026 &     5.000 &    2 &$ B_2 $&   1 &    2 &   0 &    0 &   0 &  \citep{89YaKaHi.SiH2}    \\
    1028.7264 &   1028.206 &   1028.692 &     0.520 &     0.034 &     1.000 &    2 &$ B_2 $&   1 &    2 &   0 &    1 &   0 &  \citep{89YaKaHi.SiH2}    \\
      75.2653 &     75.124 &     75.248 &     0.142 &     0.018 &     5.000 &    3 &$ A_1 $&   2 &    2 &   0 &    0 &   0 &  \citep{89YaKaHi.SiH2}    \\
    1075.2480 &   1074.682 &   1075.243 &     0.566 &     0.005 &     1.000 &    3 &$ A_1 $&   2 &    2 &   0 &    1 &   0 &  \citep{89YaKaHi.SiH2}    \\
      55.8620 &     55.680 &     55.804 &     0.182 &     0.058 &     5.000 &    3 &$ A_2 $&   1 &    3 &   0 &    0 &   0 &  \citep{89YaKaHi.SiH2}    \\
      90.1214 &     90.206 &     90.111 &    -0.085 &     0.010 &     5.000 &    3 &$ A_2 $&   3 &    1 &   0 &    0 &   0 &  \citep{89YaKaHi.SiH2}    \\
    1054.5509 &   1053.935 &   1054.484 &     0.616 &     0.067 &     1.000 &    3 &$ A_2 $&   1 &    3 &   0 &    1 &   0 &  \citep{89YaKaHi.SiH2}    \\
    1091.3028 &   1090.985 &   1091.377 &     0.317 &    -0.074 &     1.000 &    3 &$ A_2 $&   3 &    1 &   0 &    1 &   0 &  \citep{89YaKaHi.SiH2}    \\
      55.8286 &     55.641 &     55.771 &     0.187 &     0.058 &     5.000 &    3 &$ B_1 $&   0 &    3 &   0 &    0 &   0 &  \citep{89YaKaHi.SiH2}    \\
      83.7307 &     83.426 &     83.750 &     0.305 &    -0.020 &     5.000 &    3 &$ B_1 $&   2 &    1 &   0 &    0 &   0 &  \citep{89YaKaHi.SiH2}    \\
    1054.5083 &   1053.885 &   1054.440 &     0.623 &     0.068 &     1.000 &    3 &$ B_1 $&   0 &    3 &   0 &    1 &   0 &  \citep{89YaKaHi.SiH2}    \\
    1084.0720 &   1083.337 &   1084.064 &     0.735 &     0.008 &     1.000 &    3 &$ B_1 $&   2 &    1 &   0 &    1 &   0 &  \citep{89YaKaHi.SiH2}    \\
      74.1789 &     73.914 &     74.172 &     0.265 &     0.007 &     5.000 &    3 &$ B_2 $&   1 &    2 &   0 &    0 &   0 &  \citep{89YaKaHi.SiH2}    \\
      91.6751 &     91.675 &     91.678 &     0.000 &    -0.003 &     5.000 &    3 &$ B_2 $&   3 &    0 &   0 &    0 &   0 &  \citep{89YaKaHi.SiH2}    \\
    1073.9511 &   1073.248 &   1073.922 &     0.703 &     0.029 &     1.000 &    3 &$ B_2 $&   1 &    2 &   0 &    1 &   0 &  \citep{89YaKaHi.SiH2}    \\
    1092.8609 &   1092.458 &   1092.926 &     0.403 &    -0.065 &     1.000 &    3 &$ B_2 $&   3 &    0 &   0 &    1 &   0 &  \citep{89YaKaHi.SiH2}    \\
      89.2630 &     88.958 &     89.158 &     0.305 &     0.105 &     5.000 &    4 &$ A_1 $&   0 &    4 &   0 &    0 &   0 &  \citep{89YaKaHi.SiH2}    \\
     132.8721 &    132.366 &    132.890 &     0.507 &    -0.018 &     5.000 &    4 &$ A_1 $&   2 &    2 &   0 &    0 &   0 &  \citep{89YaKaHi.SiH2}    \\
     153.7492 &    153.848 &    153.744 &    -0.099 &     0.006 &     5.000 &    4 &$ A_1 $&   4 &    0 &   0 &    0 &   0 &  \citep{89YaKaHi.SiH2}    \\
    1087.7457 &   1087.013 &   1087.634 &     0.733 &     0.111 &     1.000 &    4 &$ A_1 $&   0 &    4 &   0 &    1 &   0 &  \citep{89YaKaHi.SiH2}    \\
    1133.9208 &   1132.984 &   1133.900 &     0.937 &     0.021 &     1.000 &    4 &$ A_1 $&   2 &    2 &   0 &    1 &   0 &  \citep{89YaKaHi.SiH2}    \\
    1156.7943 &   1156.528 &   1156.931 &     0.266 &    -0.137 &     1.000 &    4 &$ A_1 $&   4 &    0 &   0 &    1 &   0 &  \citep{89YaKaHi.SiH2}    \\
     115.6727 &    115.331 &    115.633 &     0.342 &     0.039 &     5.000 &    4 &$ A_2 $&   1 &    3 &   0 &    0 &   0 &  \citep{89YaKaHi.SiH2}    \\
     142.4629 &    142.020 &    142.497 &     0.443 &    -0.034 &     5.000 &    4 &$ A_2 $&   3 &    1 &   0 &    0 &   0 &  \citep{89YaKaHi.SiH2}    \\
    1115.8104 &   1115.042 &   1115.768 &     0.769 &     0.043 &     1.000 &    4 &$ A_2 $&   1 &    3 &   0 &    1 &   0 &  \citep{89YaKaHi.SiH2}    \\
    1144.1973 &   1143.347 &   1144.215 &     0.851 &    -0.018 &     1.000 &    4 &$ A_2 $&   3 &    1 &   0 &    1 &   0 &  \citep{89YaKaHi.SiH2}    \\
     115.8995 &    115.597 &    115.856 &     0.303 &     0.043 &     5.000 &    4 &$ B_1 $&   2 &    3 &   0 &    0 &   0 &  \citep{89YaKaHi.SiH2}    \\
     152.9005 &    153.073 &    152.884 &    -0.173 &     0.017 &     5.000 &    4 &$ B_1 $&   4 &    1 &   0 &    0 &   0 &  \citep{89YaKaHi.SiH2}    \\
    1116.0978 &   1115.375 &   1116.062 &     0.722 &     0.035 &     1.000 &    4 &$ B_1 $&   2 &    3 &   0 &    1 &   0 &  \citep{89YaKaHi.SiH2}    \\
    1155.9731 &   1155.777 &   1156.119 &     0.196 &    -0.145 &     1.000 &    4 &$ B_1 $&   4 &    1 &   0 &    1 &   0 &  \citep{89YaKaHi.SiH2}    \\
      89.2674 &     88.964 &     89.162 &     0.304 &     0.105 &     5.000 &    4 &$ B_2 $&   1 &    4 &   0 &    0 &   0 &  \citep{89YaKaHi.SiH2}    \\
     135.8151 &    135.614 &    135.806 &     0.201 &     0.009 &     5.000 &    4 &$ B_2 $&   3 &    2 &   0 &    0 &   0 &  \citep{89YaKaHi.SiH2}    \\
    1087.7515 &   1087.020 &   1087.640 &     0.732 &     0.111 &     1.000 &    4 &$ B_2 $&   1 &    4 &   0 &    1 &   0 &  \citep{89YaKaHi.SiH2}    \\
    1137.3989 &   1136.796 &   1137.437 &     0.603 &    -0.038 &     1.000 &    4 &$ B_2 $&   3 &    2 &   0 &    1 &   0 &  \citep{89YaKaHi.SiH2}    \\
     164.2316 &    163.770 &    164.147 &     0.462 &     0.084 &     5.000 &    5 &$ A_1 $&   2 &    4 &   0 &    0 &   0 &  \citep{89YaKaHi.SiH2}    \\
     211.6531 &    211.413 &    211.654 &     0.240 &     0.000 &     5.000 &    5 &$ A_1 $&   4 &    2 &   0 &    0 &   0 &  \citep{89YaKaHi.SiH2}    \\
    1164.5892 &   1163.715 &   1164.517 &     0.874 &     0.073 &     1.000 &    5 &$ A_1 $&   2 &    4 &   0 &    1 &   0 &  \citep{89YaKaHi.SiH2}    \\
    1215.2686 &   1214.656 &   1215.365 &     0.613 &    -0.097 &     1.000 &    5 &$ A_1 $&   4 &    2 &   0 &    1 &   0 &  \citep{89YaKaHi.SiH2}    \\
     130.0721 &    129.618 &    129.906 &     0.454 &     0.166 &     5.000 &    5 &$ A_2 $&   1 &    5 &   0 &    0 &   0 &  \citep{89YaKaHi.SiH2}    \\
     190.7850 &    190.336 &    190.760 &     0.449 &     0.025 &     5.000 &    5 &$ A_2 $&   3 &    3 &   0 &    0 &   0 &  \citep{89YaKaHi.SiH2}    \\
     231.4354 &    231.741 &    231.410 &    -0.305 &     0.025 &     5.000 &    5 &$ A_2 $&   5 &    1 &   0 &    0 &   0 &  \citep{89YaKaHi.SiH2}    \\
    1128.2434 &   1127.370 &   1128.075 &     0.874 &     0.169 &     1.000 &    5 &$ A_2 $&   1 &    5 &   0 &    1 &   0 &  \citep{89YaKaHi.SiH2}    \\
    1192.8474 &   1192.000 &   1192.853 &     0.847 &    -0.006 &     1.000 &    5 &$ A_2 $&   3 &    3 &   0 &    1 &   0 &  \citep{89YaKaHi.SiH2}    \\
    1236.8763 &   1236.852 &   1237.110 &     0.024 &    -0.234 &     1.000 &    5 &$ A_2 $&   5 &    1 &   0 &    1 &   0 &  \citep{89YaKaHi.SiH2}    \\
     130.0715 &    129.617 &    129.906 &     0.454 &     0.166 &     5.000 &    5 &$ B_1 $&   0 &    5 &   0 &    0 &   0 &  \citep{89YaKaHi.SiH2}    \\
     189.9302 &    189.340 &    189.919 &     0.590 &     0.012 &     5.000 &    5 &$ B_1 $&   2 &    3 &   0 &    0 &   0 &  \citep{89YaKaHi.SiH2}    \\
     216.3755 &    215.847 &    216.421 &     0.529 &    -0.045 &     5.000 &    5 &$ B_1 $&   4 &    1 &   0 &    0 &   0 &  \citep{89YaKaHi.SiH2}    \\
    1128.2426 &   1127.369 &   1128.074 &     0.874 &     0.169 &     1.000 &    5 &$ B_1 $&   0 &    5 &   0 &    1 &   0 &  \citep{89YaKaHi.SiH2}    \\
    1191.7737 &   1190.762 &   1191.752 &     1.012 &     0.021 &     1.000 &    5 &$ B_1 $&   2 &    3 &   0 &    1 &   0 &  \citep{89YaKaHi.SiH2}    \\
    1219.9643 &   1219.063 &   1220.030 &     0.901 &    -0.066 &     1.000 &    5 &$ B_1 $&   4 &    1 &   0 &    1 &   0 &  \citep{89YaKaHi.SiH2}    \\
     164.1932 &    163.723 &    164.110 &     0.470 &     0.083 &     5.000 &    5 &$ B_2 $&   1 &    4 &   0 &    0 &   0 &  \citep{89YaKaHi.SiH2}    \\
     205.7064 &    204.907 &    205.755 &     0.799 &    -0.048 &     5.000 &    5 &$ B_2 $&   3 &    2 &   0 &    0 &   0 &  \citep{89YaKaHi.SiH2}    \\
     231.8532 &    232.109 &    231.835 &    -0.256 &     0.018 &     5.000 &    5 &$ B_2 $&   5 &    0 &   0 &    0 &   0 &  \citep{89YaKaHi.SiH2}    \\
    1164.5381 &   1163.653 &   1164.464 &     0.885 &     0.074 &     1.000 &    5 &$ B_2 $&   1 &    4 &   0 &    1 &   0 &  \citep{89YaKaHi.SiH2}    \\
    1208.3132 &   1207.099 &   1208.305 &     1.214 &     0.009 &     1.000 &    5 &$ B_2 $&   3 &    2 &   0 &    1 &   0 &  \citep{89YaKaHi.SiH2}    \\
    1237.2663 &   1237.196 &   1237.494 &     0.070 &    -0.228 &     1.000 &    5 &$ B_2 $&   5 &    0 &   0 &    1 &   0 &  \citep{89YaKaHi.SiH2}    \\
\hline\hline
\end{longtable}
}

During the refinement we noticed an inconsistency in the experimental  energy levels from Ref.~\citep{02IsMuMi.SiH2}. For example, their  term value 1014.14~\cm\ (0,1,0), $J_{K_a,K_c} = 1_{11}$ can be compared to a more accurate value 1010.6389~\cm\ from Ref.~\citep{89YaKaHi.SiH2}. Here $K_a$ and $K_c$ are the asymmetric top quantum numbers representing the projections of $J$ along the principal axes $a$ and $c$, respectively.
In fact, after a preliminary fitting to the high resolution data from Refs.~\citep{89YaKaHi.SiH2,99HiIsxx.SiH2,91IsKaxx.SiH2} we noticed a systematic shift of around 3.5~\cm\ for all term values from Ref.~\citep{02IsMuMi.SiH2}, with the only explanation being miscalibrated experimental energies. We therefore corrected all these term values from Ref.~\citep{02IsMuMi.SiH2} by a constant shift of -3.5~\cm\ (the values in Table~\ref{t:obs-calc} have been shifted). This improved the quality of the refinement immediately.

In the refinement, the ro-vibrational eigenfunctions of the Hamiltonian constructed with the \ai\ potential $V$ act as a basis set for solving the `perturbed' ro-vibrational Hamiltonian with the refined potential $V^{\p}=V+\Delta V$. The latter is represented using the same expansion as in Eq.~\eqref{e:PES} so the refined parameters $f^{\p}_{ijk}=f_{ijk}+\Delta f_{ijk}$, where the corrections $\Delta f_{ijk}$ are determined in the fitting. The stability of the refinement is controlled by simultaneously fitting to the original \ai\ dataset~\citep{03YuCaJe.PH3}, ensuring that the shape of the PES remains reasonable.

\begin{figure}[ht]
\centering
\includegraphics[width=0.8\columnwidth]{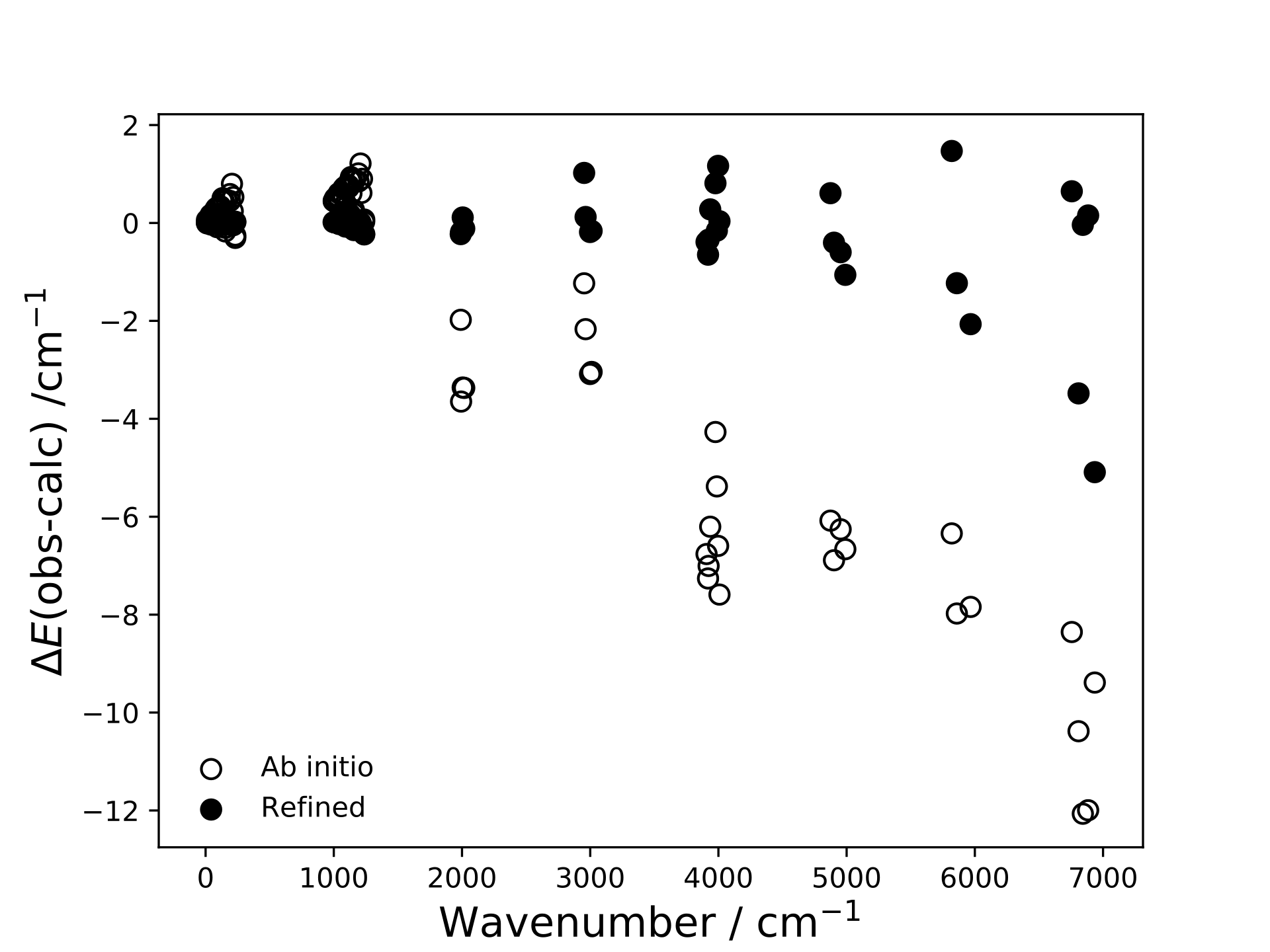}
\caption{Residuals (observed-calculated) for the \ai\ (empty circles) and the refined (full circles) term values from Table~\ref{t:obs-calc}.}
\label{fig:residuals}
\end{figure}

Only 7 expansion parameters were varied in the refinement: the linear parameters $f_{100}$, $f_{001}$; the quadratic parameters $f_{200}$, $f_{101}$, $f_{110}$, $f_{002}$; and one cubic parameter $f_{003}$. The quality of the final PES refinement is detailed in Table~\ref{t:obs-calc} and illustrated in Fig.~\ref{fig:residuals}. The different weights used in the refinement reflect the experimental uncertainty. Also shown in Table~\ref{t:obs-calc} are the variationally computed energy levels using the \ai\ and refined PESs and the corresponding residuals (observed$-$calculated), all in \cm.  From Fig.~1 we can clearly see that the refined PES calculations have far smaller residuals compared to the \ai\ results and that the accuracy of the computed term values has improved. The 100 energy levels used in the refinement were reproduced with an unweighted rms error of 0.73~\cm\ compared to the rms error of 3.43~\cm\ for the \ai\ PES. The rotational and $\nu_2$ term values from Ref.~\citep{89YaKaHi.SiH2} were reproduced with an rms error of 0.074~\cm, the vibrational band centers reported in Ref.~\citep{91IsKaxx.SiH2} were reproduced with an rms error of 0.73~\cm\ and the corrected ro-vibrational $J_{K_a,K_c}$ term values from Ref.~\citep{02IsMuMi.SiH2} were reproduced with an rms error of 1.45~\cm.

The equilibrium geometry was refined to the ground state rotational energy levels up to $J=5$, yielding the values $r_{\rm e} = 1.51440$~\AA\ and $\alpha_{\rm e} =92.005^{\circ}$. These are almost identical to the experimental values $r_{\rm exp} = 1.5137\,\pm\,0.0003$~\AA\ and $\alpha_{\rm exp} =92.04\,\pm\,0.05^{\circ}$, derived in a combined analysis of high-resolution spectroscopic data of SiH$_2$, SiHD and SiD$_2$~\citep{16KoMaSt.SiH2}. Both the \ai\ and refined PESs of SiH$_2$ are provided as part of the supplementary material along with a Fortran routine to construct the surfaces.

\section{The CATS line list}
\label{sec:results}

\begin{table}
\caption{\label{tab:states-file}Extract from the \texttt{.states} file of the \lml\ line list}
\centering
{\tt\tiny
\begin{tabular}[h!]{crccccrrcccccrrcc}
\hline\hline
\multicolumn{1}{c}{$i$} & \multicolumn{1}{c}{$\tilde{E}_i$} & $g_i$ & $J$ & $\Gamma_{\textrm{tot}}$ & $v_1$ & $v_2$ & $v_3$ & $\Gamma_{\textrm{vib}}$ & $K_a$  & $K_c$ & $\tau_{\textrm{rot}}$ & $\Gamma_{\textrm{rot}}$ & $C_i$  & $n_1$ & $n_2$ & $n_3$ \\
\hline
        1527& 19978.952397  &    3    &   1& A2  &   2&   8 &  5& B2 &    0 &  1& B1&    0.98 &  2 &  5&   8 \\
        1528& 19982.538080  &    3    &   1& A2  &   0&  20 &  1& B2 &    0 &  1& B1&    0.98 &  0 &  1&  20 \\
        1529& 19986.504561  &    3    &   1& A2  &  10&   2 &  1& B2 &    0 &  1& B1&   -0.99 &  0 & 11&   2 \\
        1530& 19988.143253  &    3    &   1& A2  &  11&   2 &  0& A1 &    1 &  1& A2&    0.99 &  0 & 11&   2 \\
        1531&    10.721260  &    9    &   1& B1  &   0&   0 &  0& A1 &    0 &  1& B1&    1.00 &  0 &  0&   0 \\
        1532&  1009.393331  &    9    &   1& B1  &   0&   1 &  0& A1 &    0 &  1& B1&   -1.00 &  0 &  0&   1 \\
        1533&  1989.081019  &    9    &   1& B1  &   0&   2 &  0& A1 &    0 &  1& B1&    1.00 &  0 &  0&   2 \\
        1534&  2004.596674  &    9    &   1& B1  &   0&   0 &  1& B2 &    1 &  1& A2&   -1.00 &  0 &  1&   0 \\
        1535&  2016.232068  &    9    &   1& B1  &   1&   0 &  0& A1 &    0 &  1& B1&    1.00 &  1 &  0&   0 \\
        1536&  2962.467155  &    9    &   1& B1  &   0&   3 &  0& A1 &    0 &  1& B1&   -1.00 &  0 &  0&   3 \\
\hline
\hline
\end{tabular}
}
{\flushleft \footnotesize
$i$:   State counting number; \\
$\tilde{E}_i$: State energy in \cm; \\
$g_i$: State degeneracy; \\
$J$: Total angular momentum quantum number; \\
$\Gamma_{\textrm{tot}}$: Overall symmetry of state in \Cv{2}(M); \\
$v_1$--$v_3$:  Vibrational (normal mode) quantum numbers; \\
$\Gamma_{\textrm{vib}}$: Vibrational symmetry in \Cv{2}(M); \\
$K_a$:  Asymmetric top quantum number;\\
$K_c$:  Asymmetric top quantum number;\\
$\tau_{\textrm{rot}}$:  Rotational parity (0 or 1); \\
$\Gamma_{\textrm{rot}}$: Rotational symmetry in \Cv{2}(M);\\
$C_{i}$: Largest coefficient  used in the \TROVE\ assignment;\\
$n_1$--$n_3$:  Vibrational (TROVE) quantum numbers. \\
}
\end{table}

\begin{table}
\caption{\label{tab:trans-file}Extract from the \texttt{.trans} file of the \lml\ line list}
{\tt
\centering
\begin{tabular}[h!]{rrr}
\hline\hline
\multicolumn{1}{c}{$f$} & \multicolumn{1}{c}{$i$} & \multicolumn{1}{c}{$A_{fi}$}   \\
\hline
      149895   &      153613 &  1.0606e-06   \\
       24599   &       22431 &  1.0490e-01    \\
      284830   &      306093 &  6.3416e-10   \\
       74616   &       84147 &  6.5034e-08    \\
      186408   &      165765 &  1.1144e-07    \\
       84280   &      100878 &  4.1022e-07    \\
      224529   &      228578 &  2.8423e-04    \\
       54609   &       46478 &  1.4376e-02    \\
      142008   &      130207 &  3.6173e-10    \\
\hline
\hline
\end{tabular}
}
{\flushleft
$f$:  Upper state ID counting number; \\
$i$: Lower state ID counting number; \\
$A_{fi}$:  Einstein $A$ coefficient in s$^{-1}$. \\
}
\end{table}

The \lml\ line list contains 254\,061\,207 transitions connecting 369\,973 ro-vibrational states and is provided in the ExoMol data format~\citep{jt631}. Extracts from the \texttt{.states} and \texttt{.trans} files are given in Tables~\ref{tab:states-file} and \ref{tab:trans-file}, respectively. The \texttt{.states} file contains all the computed ro-vibrational energy levels (in cm$^{-1}$). Each level has a unique state counting number, symmetry and quantum number labelling and the contribution $C_i$ from the largest eigen-coefficient used to assign the ro-vibrational state in \TROVE. The \texttt{.trans} files are split into ten 1000~cm$^{-1}$ wavenumber windows and contain all the computed transitions with upper and lower state ID labels and Einstein $A$ coefficients.

The normal mode quantum numbers $v_1$--$v_3$ were first reconstructed for $J=0$ and then propagated to all ro-vibrational states. These are related to the \TROVE\ (local mode) vibrational quantum numbers $n_1$--$n_3$ as follows:
\begin{eqnarray}
  v_1 +v_3 &=& n_1 + n_2 \quad ({\rm stretching}), \\
  v_3 \; ({\rm odd})  \quad  & \leftrightarrow & \quad (n_1,n_2) B_2, \\
  v_2 &=& n_3   \quad ({\rm bending}).
\end{eqnarray}
In correlating the normal mode $(v_1,v_3)$ and local mode $(n_1,n_2)$ pairs we also assumed that the energy increases with $v_3$.
The asymmetric top quantum number $K_a$ coincides with the TROVE rotational quantum number $K$. The corresponding quantum number $K_c$ was obtained using the symmetry properties of the oblate rotor \citep{98BuJexx}, see Table~\ref{tab:Ka:Kc}.  Because of the way \TROVE\ builds the symmetrized ro-vibrational basis set~\citep{17YuYaOv.methods}, the connection between the assignment and the primitive basis functions is not always straightforward and in cases of very small values of $|C_i|$, the assignment should be regarded as indicative.

\begin{table}
\caption{\label{tab:Ka:Kc} Symmetry properties of the $J_{K_a,K_c}$ states of SiH$_2$ in the \Cv{2}(M) symmetry group \citep{98BuJexx}.}
\centering
\begin{tabular}[h!]{ccccccc}
\hline
$K_a$ & $K_c$ & $\Gamma_{\rm rot}$ && $K_a$ & $K_c$ & $\Gamma_{\rm rot}$ \\
\hline
Even &  Even & $A_1$ && Odd &  Even & $B_2$ \\
Even &  Odd  & $B_1$ && Odd &  Odd & $A_2$ \\
\hline
\end{tabular}
\end{table}

\begin{figure}
\includegraphics[width=0.5\columnwidth]{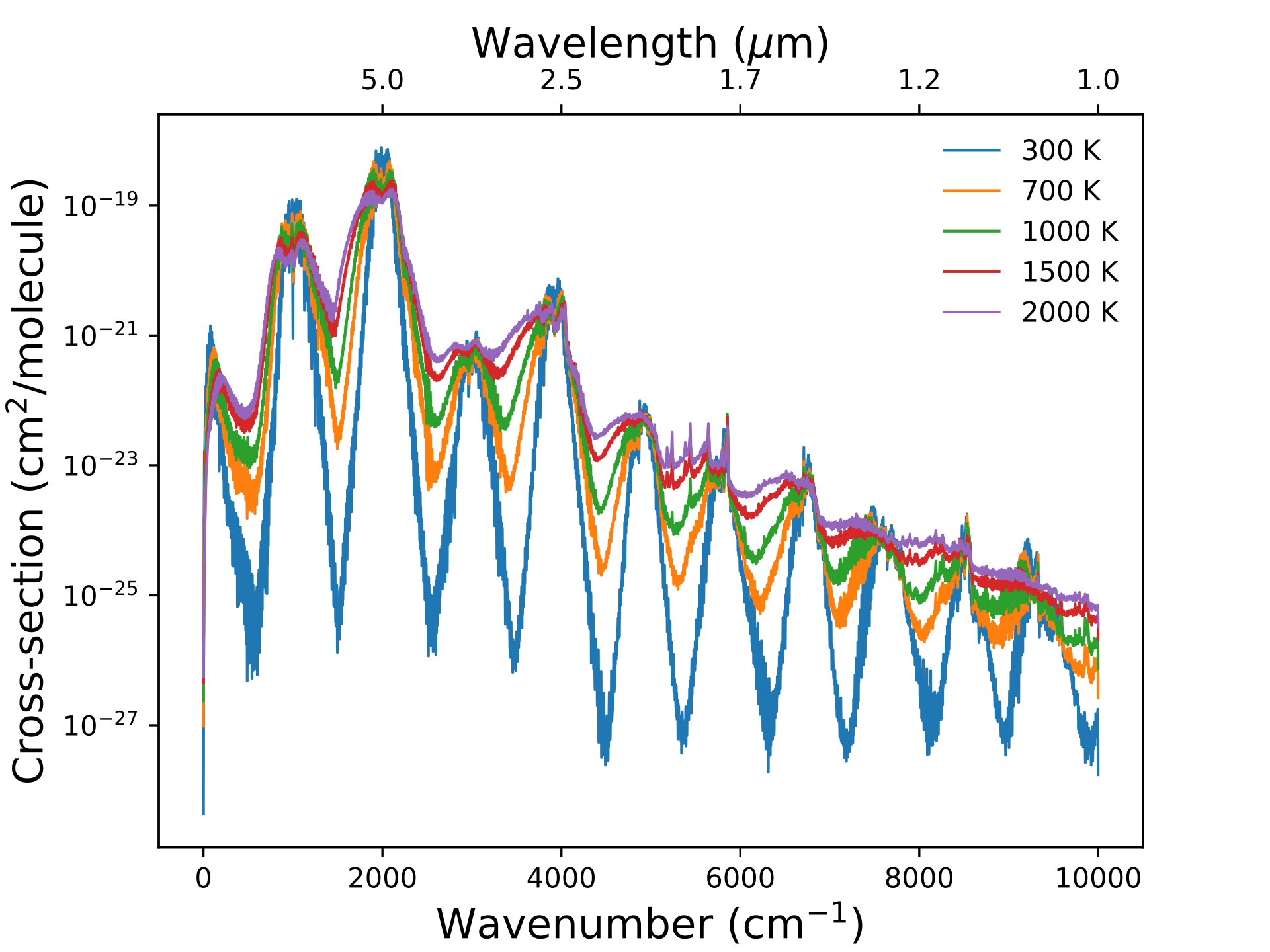}
\includegraphics[width=0.5\columnwidth]{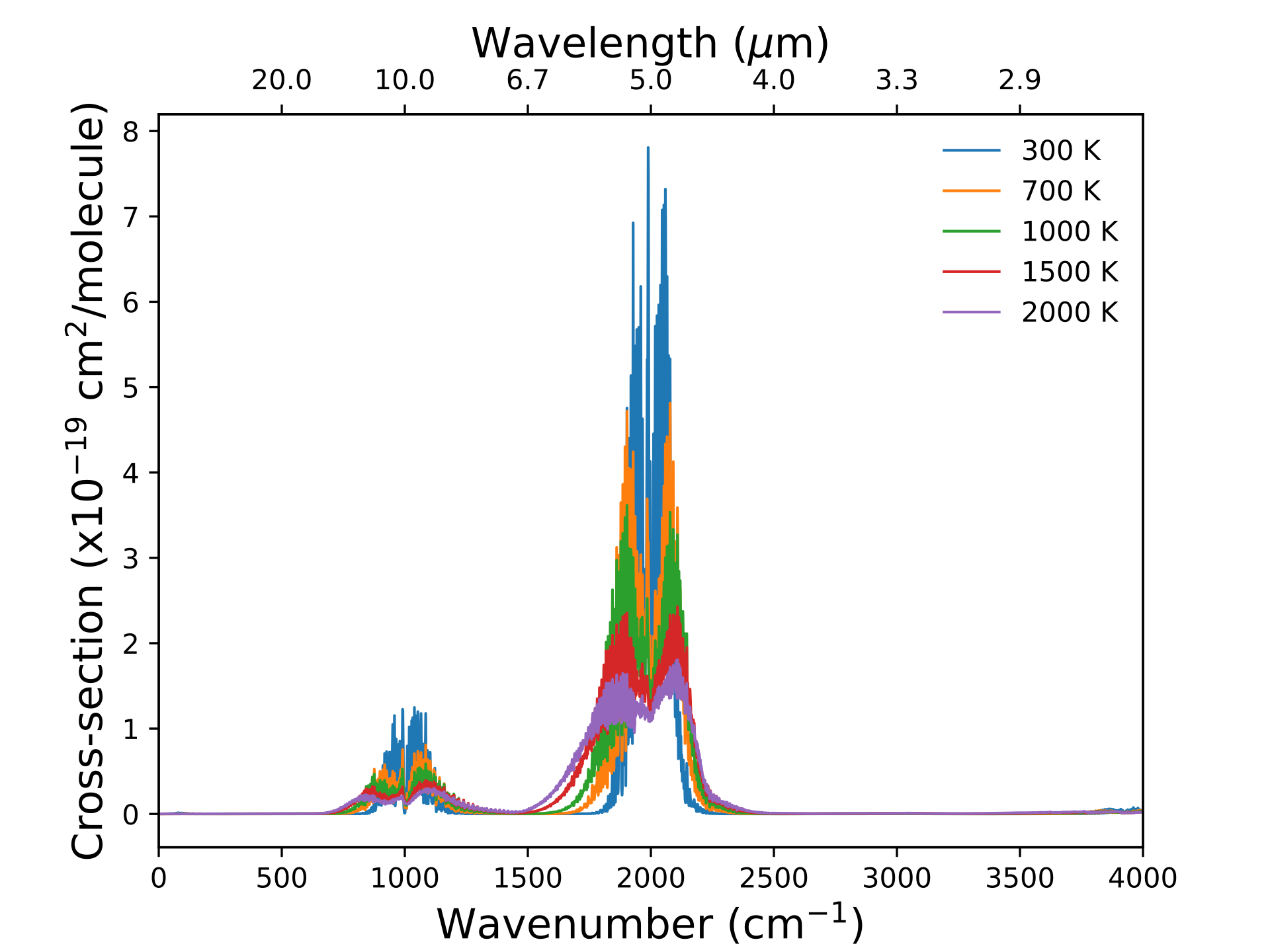}
\caption{Absorption cross-sections of \sihh\ for various temperatures. Left:  0--10\,000~\cm\ range (logarithmic scale); Right:
0--4000~\cm\ range (linear scale).  Spectral simulations used a Gaussian line profile with HWHM of 1~\cm\ at a resolution of 1~\cm.}
\label{cross-sections-300-2000K}
\end{figure}

\begin{figure}
\centering
\includegraphics[width=0.9\columnwidth]{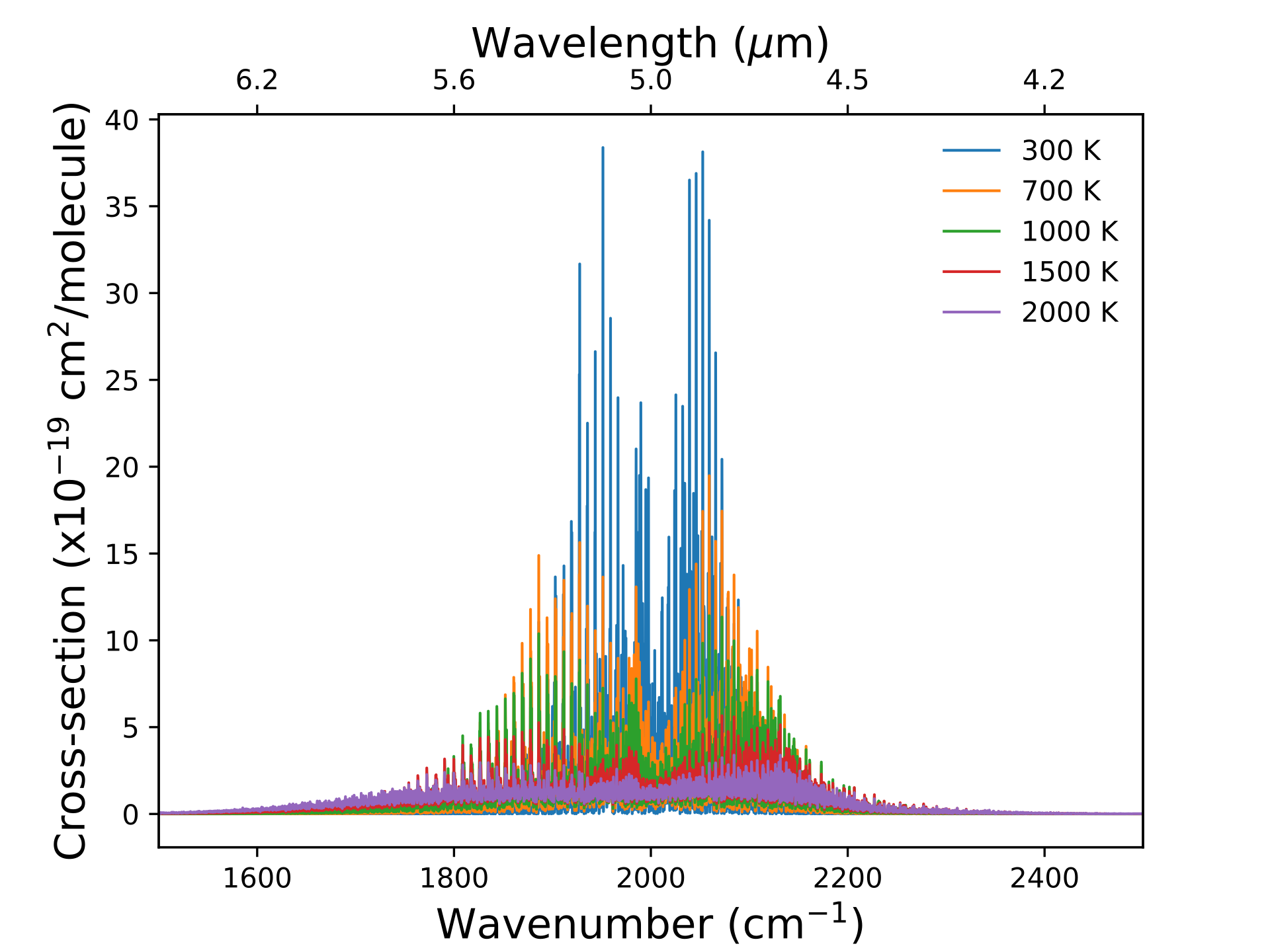}
\caption{Absorption cross-sections of \sihh\ for various temperatures, focusing on the $\nu_1$ band. Spectral simulations used a Gaussian line profile with HWHM of 0.1~\cm\ at a resolution of 0.1~\cm. }
\label{cross-sections-zoom-300-2000K-linear}
\end{figure}

To illustrate the \lml\ line list, absorption cross-sections have been generated at 300, 700, 1000, 1500 and 2000~K and the results are shown in Figures~\ref{cross-sections-300-2000K} and \ref{cross-sections-zoom-300-2000K-linear}. Spectral simulations used a Gaussian line profile and were performed with the ExoCross program~\citep{ExoCross}. In Fig.~\ref{cross-sections-300-2000K}, the entire computed spectrum of \sihh\ is displayed at logarithmic scale and the linear scale is displayed up to 4000 \icm. The strongest bands of SiH$_2$ are 10~$\mu$m ($\nu_2$) and 5~$\mu$m ($2\nu_2$, $\nu_1$ and $\nu_3$), followed by the 2.5~$\mu$m ($2\nu_1$ and $2\nu_3$) bands, which agrees with the ro-vibrational spectrum from Ref.~\cite{93GaRoYa.SiH2}. The pure rotational band is weak owing to the small equilibrium dipole moment of SiH$_2$. The corresponding values of the transition dipole moments are listed in Table~\ref{tab:trans:dipole}.
As the temperature increases, weaker band features become more prominent and the spectrum flattens. This can be seen in closer detail in Fig.~\ref{cross-sections-300-2000K} and Fig.~\ref{cross-sections-zoom-300-2000K-linear}.

\begin{table}
\caption{\label{tab:trans:dipole} Vibrational transition dipole moments (TDM) for the strongest absorption bands of SiH$_2$ (Debye). The band centers (BC)  are theoretical (TROVE) values (\icm).}
\centering
\begin{tabular}{lrr}
\hline\hline
Band              &    BC (\icm)   & TDM  (Debye)  \\
\hline
  g.s.             &     0.00   &     0.1093  \\
 $ \nu_2          $&   998.61   &     0.1752  \\
 $ 2\nu_2         $&  1978.35   &     0.1352  \\
 $ \nu_3          $&  1993.00   &     0.2240  \\
 $ \nu_1          $&  2005.54   &     0.1444  \\
 $ 3\nu_2         $&  2951.68   &     0.0029  \\
 $ \nu_2+\nu_3    $&  2974.73   &     0.0036  \\
 $ \nu_1+\nu_2    $&  2998.78   &     0.0076  \\
 $ 2\nu_3         $&  3907.79   &     0.0102  \\
 $ \nu_1+\nu_3    $&  3913.39   &     0.0145  \\
 $ 2\nu_1         $&  3923.64   &     0.0092  \\
 $ 2\nu_2+\nu_3   $&  3952.47   &     0.0055  \\
\hline\hline
\end{tabular}
\end{table}

The first excited triplet state ($\tilde{a}^3B_1$) of \sihh\ is around 7000~\cm~\citep{87BeGrCh.SiH2} while the first excited singlet state (\ab) lies at 15\,500~\cm~\citep{98EsCaxx.SiH2}. The presence of these states would have had an effect on the accuracy of the computed \ai\ PES and DMS,
due to Renner-Teller and spin-orbit interactions between these electronic states, which were not taken into account here.  Furthermore, the PES was only refined to term values below 7000~\cm. Use of the \lml\ line list above 7000~\cm, or for transitions originating from states above 7000~\cm, should therefore be treated with a degree of caution. That said, the associated transition intensities will be very weak and this should not overly affect the quality of the line list. 

\subsection{Partition function of silylene}

\begin{figure}
\centering
\includegraphics[width=0.9\columnwidth]{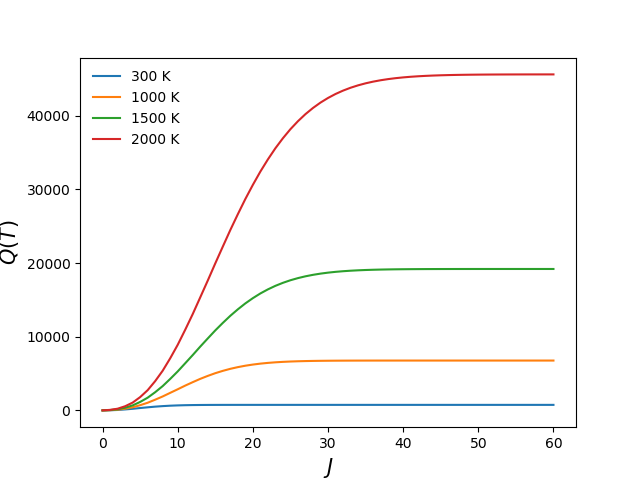}
\caption{Convergence of the temperature-dependent partition function $Q(T)$ of \sihh\ against the rotational angular momentum quantum number $J$ at 300, 1000, 1500 and 2000~K.}
\label{pfn-convergence-300-2000K}
\end{figure}

The temperature-dependent partition function,
\begin{equation}
\label{eq:pfn}
Q(T)=\sum_{i} g_i \exp\left(-\frac{E_i}{kT}\right) ,
\end{equation}
where $g_i=g_i^{(\rm ns)}(2J_i+1)$ is the degeneracy of a state $i$ with energy $E_i$ and rotational quantum number $J_i$, has been evaluated at 1~K intervals in the range 0--2000~K. The convergence of $Q(T)$ is illustrated in Fig.~\ref{pfn-convergence-300-2000K} for $T$ = 300, 1000, 1500 and 2000~K.  The flattening of all four lines indicates that convergence has been achieved after approximately $J$ = 10, 30, 40 and 50, respectively.

\section{Rotational energy level clustering}
\label{s:cluster}

The equilibrium interbond angle of SiH$_2$ is close to 90$^\circ$ with the masses of the hydrogen atoms significantly smaller than that of Si. These properties are known to lead to so-called rotational cluster states~\citep{97JeOsKo.cluster}, where a group of
rotational energy levels become quasi-degenerate at high rotational excitation. Dorney and Watson~\citep{72DoWaxx.cluster} were the first to explain cluster formation in terms of classical rotation about symmetrically equivalent axes associated with ``stable'' axes of rotation, about which the molecule prefers to rotate. Subsequently, Harter and co-workers (see, for example, Refs.~\citep{84HaPaxx.cluster,88Harter.cluster}) developed classical models for the description of cluster formation in XY$_{N}$ molecules and introduced the concept of a rotational energy surface (RES), which defines the pattern of the rotational energy levels. The clustering of two or four quasi-degenerate energy levels corresponds classically to the appearance of two or four stable rotation axes, where clockwise and anticlockwise rotations are regarded independently. The two-fold degeneracy is the well known $K$-type doubling~\citep{82PaAlxx}, while four-fold cluster states in XH$_2$-type molecules were predicted~\citep{88ZhPaxx.cluster,91Lehmann.cluster,93KoKlJe.cluster,94KoJexx.cluster,96KoPaxx.cluster,96KoJePo.cluster,97GoPaJe.cluster,00Jensen.cluster} and first observed in the H$_2$Se molecule~\citep{93KoJexx.cluster,93JeKozz.cluster,95FlCaBu.cluster}.

\begin{figure}
\centering
\includegraphics[width=0.8\columnwidth]{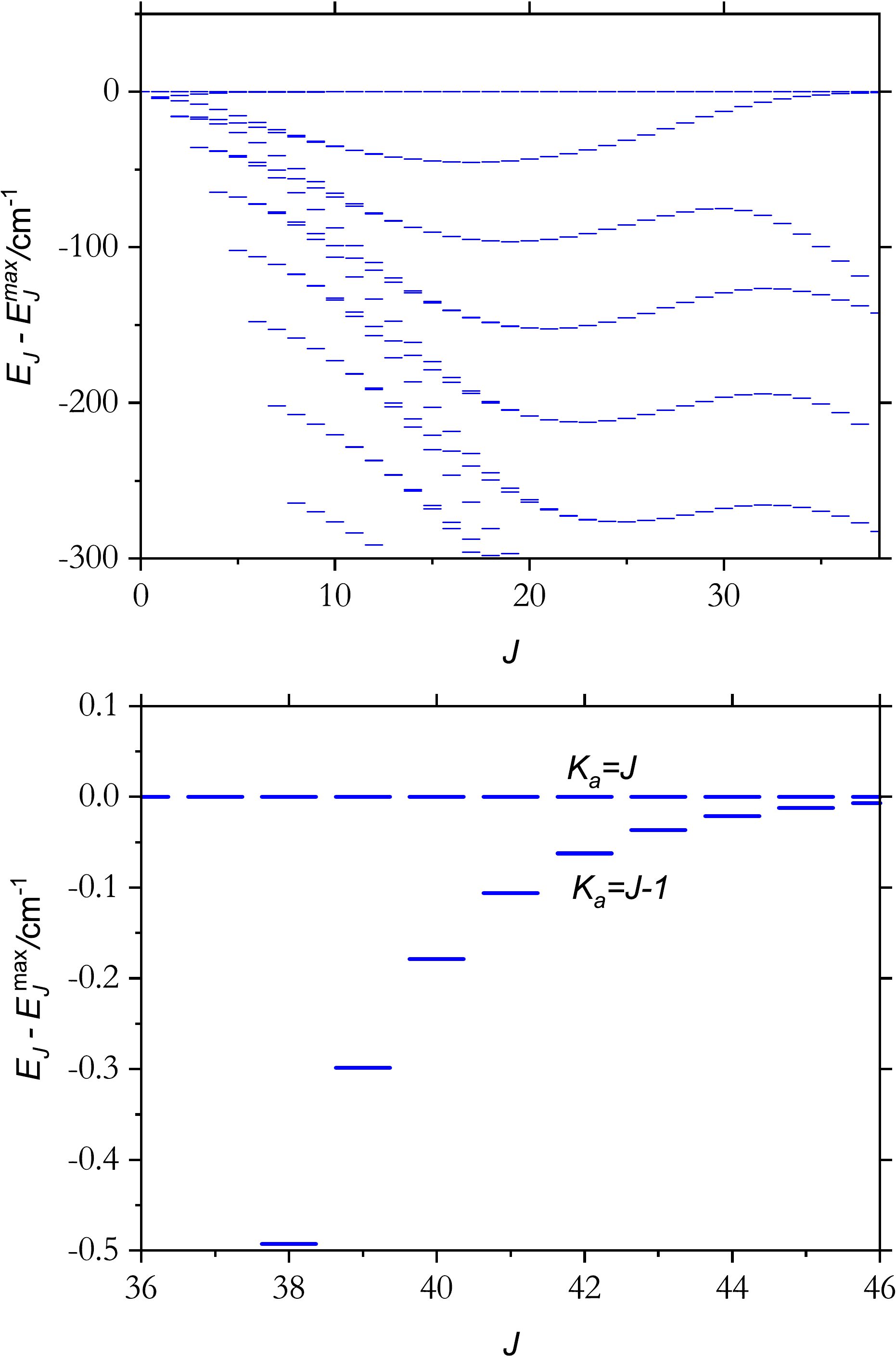}
\caption{Rotational energy level clustering in the ground vibrational state of SiH$_2$. In the top panel, the energy difference $\tilde{E}_{J,i} - \tilde{E}_{J,i}^{\rm max}$, see Eq.~\eqref{eq:en_diff}, has been plotted for each rotational state, with energy $\tilde{E}_{J,i}$ relative to the maximum energy $\tilde{E}_{J,i}^{\rm max}$ in its $J$ multiplet. After a certain critical $J$ value the energy difference begins to decrease leading to the formation of four-fold rotational cluster states. A close up of the first cluster state being formed is shown in the bottom panel.}
\label{fig:cluster}
\end{figure}

In SiH$_2$, we have found that the rotational energies display typical four-fold clustering behaviour, illustrated in Fig.~\ref{fig:cluster}, where we have plotted the reduced rotational energy difference,
\begin{equation}
\label{eq:en_diff}
\tilde{E}_{J,i}^{\rm red}  = \tilde{E}_{J,i} - \tilde{E}_{J,i}^{\rm max}.
\end{equation}
Here, $\tilde{E}_{J,i}^{\rm max}$ is the maximal energy level for a given $J$-manifold, which for SiH$_2$ corresponds to $K_a=J$. As $J$ increases the separation between the two pairs of levels with $K_a=J$ ($K_c=0$ and $K_c=1$)  and $K_a = J-1$ ($K_c=1$ and $K_c=2$) increases, which is common in asymmetric-top molecules.  However, after a critical $J$ value of $J_{\rm cr} \approx 17$, this separation starts decreasing until four-fold cluster states form. For example, at $J=40$ in the top cluster (shown in the bottom panel of Fig.~\ref{fig:cluster}), the separation between the two pairs is only 0.2~\cm\ but at $J=50$ this gap is reduced to 0.006~\cm. The pattern of the rotational energy levels in Fig.~\ref{fig:cluster} is typical of most XH$_2$-type molecules~\citep{88PaZhxx.cluster,93JeKozz.cluster,94KoJexx.cluster,96KoJePo.cluster,97GoPaJe.cluster}.

As mentioned above, the four-fold clusters at high $J$ can be associated with the formation of stable rotation axes around the molecular bonds Si--H$_1$ and Si--H$_2$ with four equivalent rotational motions: clockwise/anticlockwise around the Si--H$_1$ and Si--H$_2$ bonds. These rotation axes are approximately at 45$^{\circ}$ relative to the principal axes of the molecule $x$ axis (along the bisector) and $z$ axis (in plane). The other stable rotational direction is around the axis associated with the $x$ axis. The rotational motion is commonly illustrated using the rotational energy surface $E_{\theta,\phi}$~\citep{84HaPaxx.cluster,88Harter.cluster}, constructed as the minimum ro-vibrational energy for each orientation of the angular momentum vector $\vec{J}$ considered in the molecule-fixed axis system, where the orientation is described by a polar angle $\theta$ and an azimuthal angle $\phi$. Here $\theta$ $\in$ $[0,\pi]$ is the angle between $\vec{J}$ and the molecule-fixed $z$ axis and $\phi$ $\in$ $[0,2\pi]$ is the angle between the $x$ axis and the projection of $\vec{J}$ in the $xy$ plane, measured in the usual positive sense. In this picture, the RES is a manifestation of the pure rotational motion of a molecule with the vibrational motion completely frozen to the corresponding optimized geometry. If the rotational motion is classically represented by trajectories on the RES, the stationary points (zero-dimensional trajectories) represent rotation about the stabilization axes.

For SiH$_2$, we have constructed the RES from a classical ro-vibrational Hamiltonian $H_{\rm rv}$ with the $xyz$ components of the quantum-mechanical angular
momentum operator $\vecb{J}$ substituted by their classical analogues,
\begin{eqnarray}\label{e:J:semiclas}
  J_x &=& \sqrt{J(J+1)} \, \sin\theta\,\cos\phi \\
  J_y &=& \sqrt{J(J+1)}  \,\sin\theta\,\sin\phi \\
  J_z &=& \sqrt{J(J+1)} \, \cos\theta\, \hbox{,}
\end{eqnarray}
with the generalized momenta $p_n$ $=$ $\partial T/\partial {\dot q}_n$ set to zero~\citep{93KoJexx.cluster}. Here, $T$ is the classical kinetic energy and the momentum $p_n$ for $n=1, 2, 3$ is conjugate to the generalized coordinate $q_n$ $\in$ $\{ r_1, r_2, \alpha\}$. The classical Hamiltonian $H_{\rm rv}$ was defined as the same Hamiltonian used in variational \TROVE\ calculations. The RES $E_J(\theta,\phi)$ is then given by \citep{05YuThPa.PH3,06YuThJe.SbH3}
\begin{equation}\label{e:RES}
    E_J(\theta,\phi) = H_{\rm rv}(J,r_i=r_i^{\rm opt},\alpha=\alpha^{\rm opt},\theta,\phi)\hbox{,}
\end{equation}
where the classical Hamiltonian function $H_{\rm rv}$ is calculated at the optimized geometries $r_i^{\rm opt},\alpha^{\rm opt}$ for each orientation of the angular momentum defined by the polar and azimuthal angles $(\theta, \phi)$. The RES was computed on a regular grid of angular points $\theta_m,\phi_m$. The bond lengths $r_1$ and $r_2$ and the bond angle $\alpha$ were optimized at each grid
point by minimizing the classical energy $E=H_{\rm rv}(r_i,\alpha_i,\theta_m,\phi_m)$.

\begin{figure}
\centering
\includegraphics[width=0.9\columnwidth]{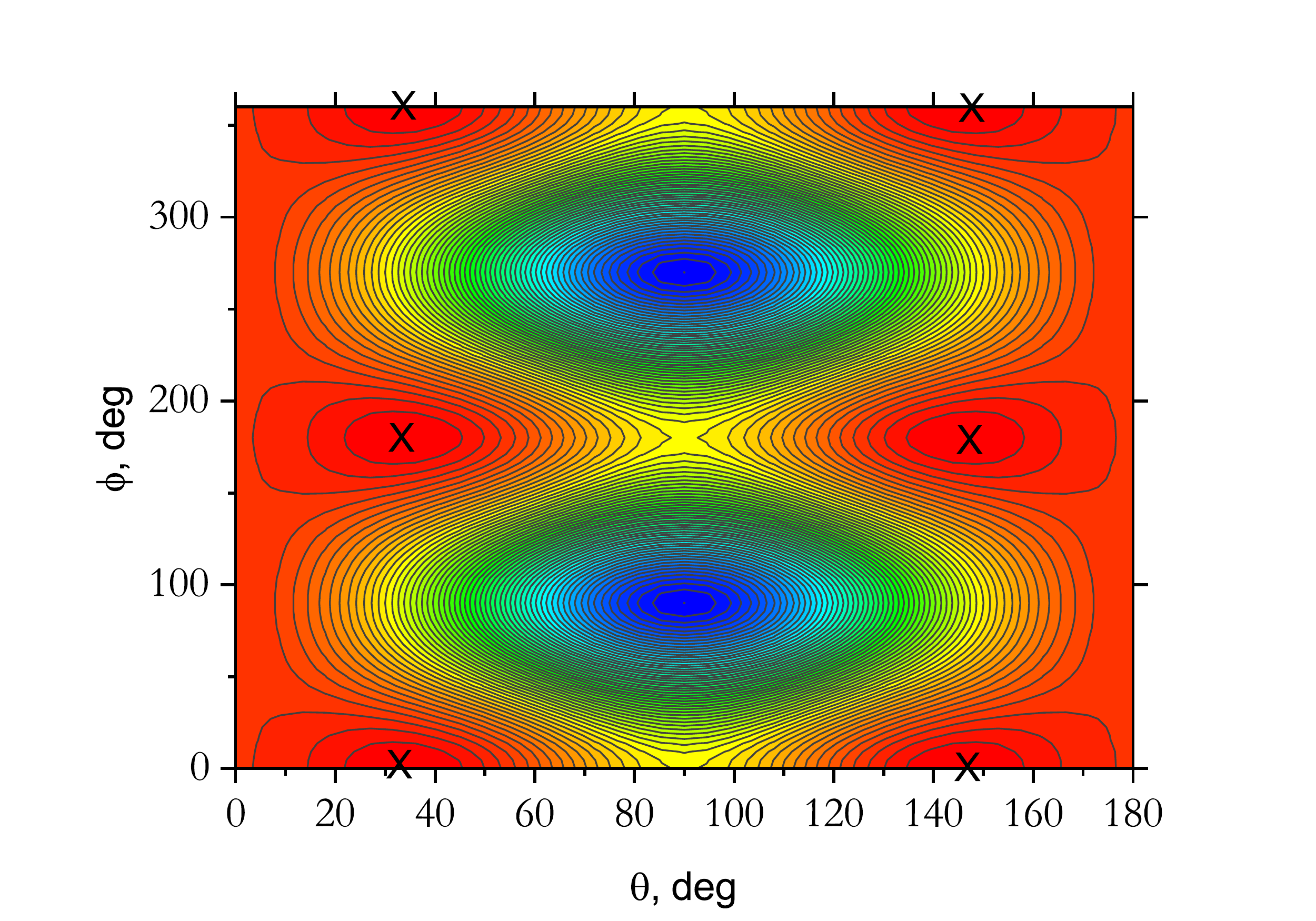}
\caption{The rotational energy surface of SiH$_2$ calculated at $J=40$. The four equivalent maxima corresponding to the four-fold rotational clusters are indicted with crosses. The colour (red to blue) represents the energy scale (high to low).}
\label{fig:RES}
\end{figure}

\begin{figure}
\centering
\includegraphics[width=0.9\columnwidth]{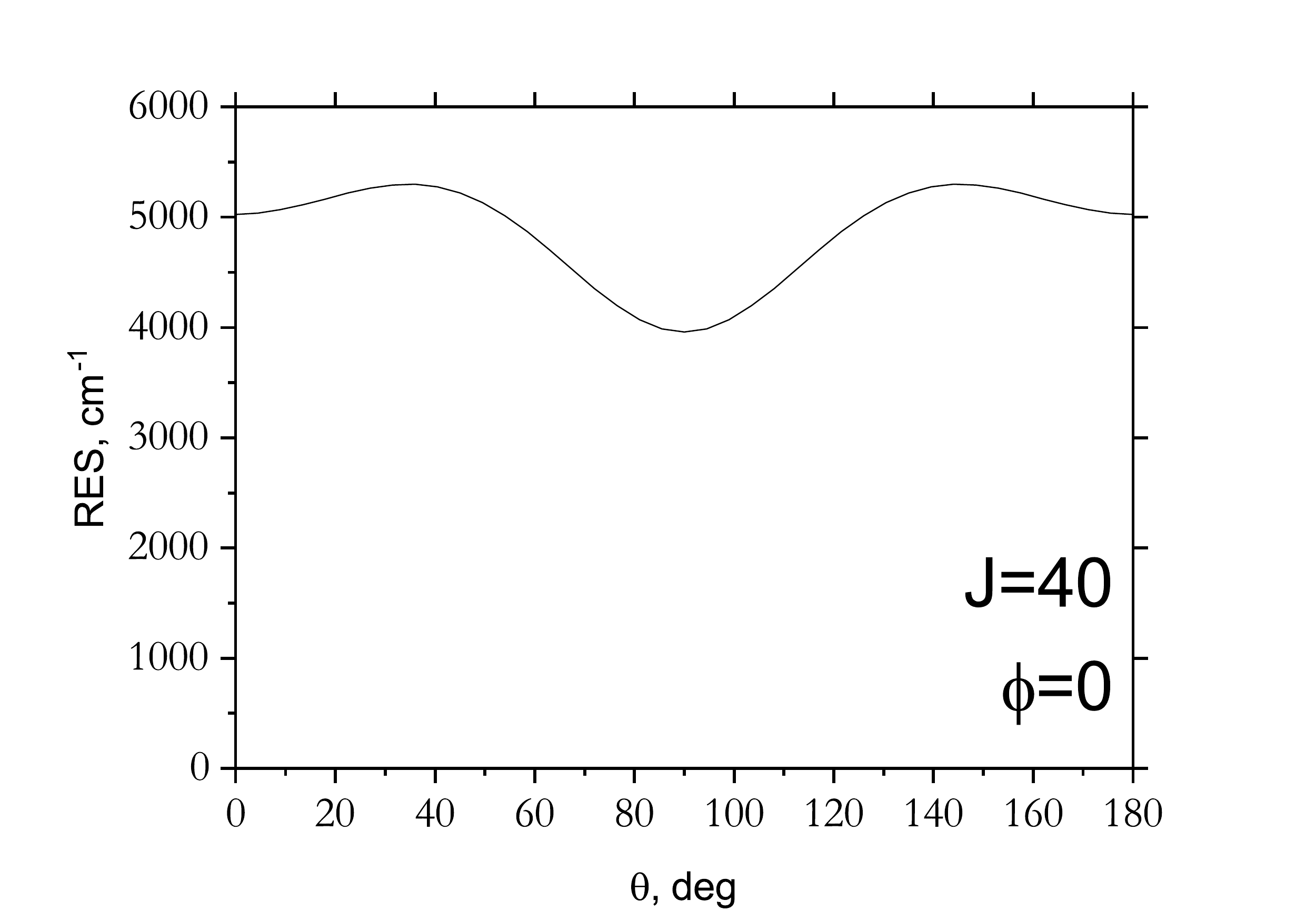}
\caption{A one-dimensional slice at $\phi=0^{\circ}$ of the rotational energy surface (RES) of SiH$_2$ calculated at $J=40$ (\icm). The two equivalent maxima are found at $\theta=36$ and $144^{\circ}$. }
\label{fig:1DRES}
\end{figure}

The RES of SiH$_2$ at $J=40$ is shown in Figs.~\ref{fig:RES} and \ref{fig:1DRES}, computed on a $40\times 80$ grid ($\theta,\phi$) of points.  The two minima ($\theta=0$ and $180^{\circ}$) and four maxima ($\theta=36^{\circ}$ and $\theta=144^{\circ}$, $\phi=0^{\circ}$ and 180$^{\circ}$ ) form the stationary points of the semi-classical description of the rotation of SiH$_2$ and can be used to interpret the rotational clustering in the quantum-mechanical description: $n$-fold degenerate energy clusters \cite{84HaPaxx.cluster,90Makarewicz.cluster} correspond to a RES with $n$ symmetrically equivalent stationary points. Coriolis-type effects break the \Cv{2} symmetry of the molecule in the cluster states. For example, the optimized geometry corresponding to the stationary point at $\theta=36^{\circ}, \phi=0$ ($J=40$) was found to be $r_1=1.519$~\AA, $r_2=1.648$~\AA\ and $\alpha=87^{\circ}$.  See also Ref.~\citep{88PaZhxx.cluster} on the bifurcation of stationary points and how it affects the formation of rotational energy clusters.

\section{Conclusion}
\label{sec:conc}

The first, comprehensive high-temperature rotation-vibration line list, named \lml, has been calculated for the electronic ground state of \sihh\ using a refined PES and a high level \ai\ DMS. The line list covers the wavenumber range 0--10\,000~\cm\ and is applicable for temperatures up to $T=2000$~K and represent an improvement for the ro-vibrational spectra of \sihh\ predicted in Refs.~\citep{93GaRoYa.SiH2,04YuBuKr.SiH2}.  The \lml\ line list is available from the ExoMol database~\citep{jt541,jt631} at \href{www.exomol.com}{www.exomol.com}  which already contains line lists for  SiH$_4$~\citep{jt701} and SiH~\citep{jt776}.
There are also recent experimental studies on SiH$_3$ \citep{19NaPiDa.SiH3}.
These line lists will facilitate spectroscopic studies on silane plasmas.
It is also hoped that our work  will stimulate more investigations into \sihh\ to enable our theoretical spectroscopic model to be benchmarked and improved upon.

We have demonstrated that SiH$_2$ forms rotational energy level clusters at high rotational excitation and established the critical $J$ value of $J_{\rm cr}\approx 17$ when the effect first becomes noticable. Cluster states are intimately linked to the phenomenon of rotationally-induced chirality in the motion of small polyatomic molecules~\citep{04BuJexx2.method} and the use of techniques from strong-field laser physics~\citep{18OwYaKu.cluster} can be utilised to produce dynamically chiral molecules through extreme rotational excitation~\citep{18OwYaYu.PH3}.

The \lml\ line list should find use in plasma physics and the monitoring of \sihh\ in different processes and reactions. Since silylene is a radical and highly reactive, whether it can accumulate in large enough quantities to be detected astronomically is debatable. As discussed above, given the complicated electronic structure of SiH$_2$ and the relatively low-lying $\tilde{a}^3B_1$ and \ab\ excited states, some caution should be exercised when using the line list above 7000~\cm, or for transitions originating from states above 7000~\cm. That said, the associated transition intensities will be very weak and we do not expect the quality of the \lml\ line list to be significantly affected.

\section*{Acknowledgements}

This work was supported by UK research councils EPSRC, under grant EP/N509577/1 and  STFC, under grant ST/R000476/1. This work made extensive use of UCL's Legion High Performance Computing facility along with the STFC DiRAC HPC facility supported by BIS National E-infrastructure capital grant ST/J005673/1 and STFC grants ST/H008586/1 and ST/K00333X/1.


\begin{thebibliography}{10}
\expandafter\ifx\csname url\endcsname\relax
  \def\url#1{\texttt{#1}}\fi
\expandafter\ifx\csname urlprefix\endcsname\relax\def\urlprefix{URL }\fi
\expandafter\ifx\csname href\endcsname\relax
  \def\href#1#2{#2} \def\path#1{#1}\fi

\bibitem{67DuHeVe.SiH2}
I.~Dubois, G.~Herzberg, R.~D. Verma, {Spectrum of SiH$_2$}, J. Chem. Phys.
  47~(10) (1967) 4262.
\newblock \href {https://doi.org/10.1063/1.1701609}
  {\path{doi:10.1063/1.1701609}}.

\bibitem{16KoMaSt.SiH2}
D.~L. Kokkin, T.~Ma, T.~Steimle, T.~J. Sears, {Detection and characterization
  of singly deuterated silylene, SiHD, via optical spectroscopy}, J. Chem.
  Phys. 144~(24) (2016) 244304.
\newblock \href {https://doi.org/10.1063/1.4954702}
  {\path{doi:10.1063/1.4954702}}.

\bibitem{87JaMeSc.SiH2}
J.~Jasinski, B.~S. Meyerson, B.~A. Scott, Mechanistic studies of chemical
  vapor-deposition, Ann. Rev. Phys. Chem. 38 (1987) 109--140.
\newblock \href {https://doi.org/10.1146/annurev.physchem.38.1.109}
  {\path{doi:10.1146/annurev.physchem.38.1.109}}.

\bibitem{90DoDoGa.SiH2}
J.~R. Doyle, D.~A. Doughty, A.~Gallagher, {Silane Dissociation Products in
  Deposition Discharges}, J. Appl. Phys. 68~(9) (1990) 4375--4384.
\newblock \href {https://doi.org/10.1063/1.346186}
  {\path{doi:10.1063/1.346186}}.

\bibitem{88JaChxx.SiH2}
J.~Jasinski, J.~O. Chu, Absolute rate constants for the reaction of silylene
  with hydrogen, silane, and disilane, J. Chem. Phys. 88 (1988) 1678--1687.
\newblock \href {https://doi.org/10.1063/1.454146}
  {\path{doi:10.1063/1.454146}}.

\bibitem{19NaPiDa.SiH3}
A.~S. Nave, A.~V. Pipa, P.~B. Davies, J.~Roepcke, J.-P.~H. van Heiden,
  {Determining a Line Strength in the $\nu_3$ Band of the Silyl Radical Using
  Quantum Cascade Laser Absorption Spectroscopy}, J. Phys. A: Math. Gen.
  123~(46) (2019) 10030--10039.
\newblock \href {https://doi.org/10.1021/acs.jpca.9b06351}
  {\path{doi:10.1021/acs.jpca.9b06351}}.

\bibitem{92Turner.SiN}
B.~E. Turner, \href{http://adsabs.harvard.edu/doi/10.1086/186324}{Detection of
  {SiN} in {IRC} + 10216}, Astrophys. J. 388 (1992) L35.
\newblock \href {https://doi.org/10.1086/186324} {\path{doi:10.1086/186324}}.
\newline\urlprefix\url{http://adsabs.harvard.edu/doi/10.1086/186324}

\bibitem{94AvBeCu.SiH2}
L.~W. Avery, M.~B. Bell, C.~T. Cunningham, P.~A. Feldman, R.~H. Hayward, J.~M.
  Macleod, H.~E. Matthews, J.~D. Wade, {Submillimeter molecular line
  observations of IRC+10216: Searches for MgH, SiH$_2$, and HCO$^+$, and
  detection of hot HCN}, Astrophys. J. 426~(2,1) (1994) 737--741.
\newblock \href {https://doi.org/10.1086/174110} {\path{doi:10.1086/174110}}.

\bibitem{99McChxx.SiH2}
D.~D.~S. MacKay, S.~B. Charnley,
  \href{https://doi.org/10.1046/j.1365-8711.1999.02175.x}{The silicon chemistry
  of {IRC}+10216}, Mon. Not. R. Astron. Soc. 302~(4) (1999) 793--800.
\newblock \href {https://doi.org/10.1046/j.1365-8711.1999.02175.x}
  {\path{doi:10.1046/j.1365-8711.1999.02175.x}}.
\newline\urlprefix\url{https://doi.org/10.1046/j.1365-8711.1999.02175.x}

\bibitem{03ApPaKa.SiH2}
Y.~Apeloig, R.~Pauncz, M.~Karni, R.~West, W.~Steiner, D.~Chapman, {Why is
  methylene a ground state triplet while silylene is a ground state singlet?},
  Organometallics 22~(16) (2003) 3250--3256.
\newblock \href {https://doi.org/10.1021/om0302591}
  {\path{doi:10.1021/om0302591}}.

\bibitem{04KaDuMa.SiH2}
A.~Kalemos, T.~H. Dunning, A.~Mavridis, {SiH$_2$, a critical study}, Mol. Phys.
  102~(23-24) (2004) 2597--2606.
\newblock \href {https://doi.org/10.1080/00268970412331293802}
  {\path{doi:10.1080/00268970412331293802}}.

\bibitem{93GaRoYa.SiH2}
W.~Gabriel, P.~Rosmus, K.~Yamashita, K.~Morokuma, P.~Palmieri, {Theoretical
  rotational vibrational-spectrum of SiH$_2$ ($X\, ^{1}A_1$ and $A\, ^3B_1$)},
  Chem. Phys. 174~(1) (1993) 45--56.
\newblock \href {https://doi.org/10.1016/0301-0104(93)80050-J}
  {\path{doi:10.1016/0301-0104(93)80050-J}}.

\bibitem{04YuBuKr.SiH2}
S.~N. Yurchenko, P.~R. Bunker, W.~P. Kraemer, P.~Jensen, {The spectrum of
  singlet SiH$_2$}, Can. J. Chem. 82~(6) (2004) 694--708.
\newblock \href {https://doi.org/10.1139/V04-030} {\path{doi:10.1139/V04-030}}.

\bibitem{05GuBuJe.SiH2}
R.~Guerout, P.~R. Bunker, P.~Jensen, W.~P. Kraemer, {A calculation of the
  rovibronic energies and spectrum of the $\tilde{B}\,^1A_1$ electronic state
  of SiH$_2$}, J. Chem. Phys. 123~(24) (2005) 244312.
\newblock \href {https://doi.org/10.1063/1.2139676}
  {\path{doi:10.1063/1.2139676}}.

\bibitem{jt654}
J.~Tennyson, {Perspective: Accurate ro-vibrational calculations on small
  molecules}, J. Chem. Phys. 145 (2016) 120901.
\newblock \href {https://doi.org/10.1063/1.4962907}
  {\path{doi:10.1063/1.4962907}}.

\bibitem{jt541}
S.~N. Yurchenko, J.~Tennyson, {ExoMol: molecular line lists for exoplanet and
  other atmospheres}, in: C.~Stehl{\'e}, C.~Jobin, L.~d'Hendercourt (Eds.),
  European Conference on Laboratory Astrophysics -- ECLA, Vol.~58 of Euro.
  Astron. Soc. Publications Series, 2012, pp. 243--248.

\bibitem{jt631}
J.~Tennyson, S.~N. Yurchenko, A.~F. Al-Refaie, E.~J. Barton, K.~L. Chubb, P.~A.
  Coles, S.~Diamantopoulou, M.~N. Gorman, C.~Hill, A.~Z. Lam, L.~Lodi, L.~K.
  McKemmish, Y.~Na, A.~Owens, O.~L. Polyansky, T.~Rivlin, C.~Sousa-Silva, D.~S.
  Underwood, A.~Yachmenev, E.~Zak, {The ExoMol database: molecular line lists
  for exoplanet and other hot atmospheres}, J. Mol. Spectrosc. 327 (2016)
  73--94.
\newblock \href {https://doi.org/10.1016/j.jms.2016.05.002}
  {\path{doi:10.1016/j.jms.2016.05.002}}.

\bibitem{jt701}
A.~Owens, S.~N. Yurchenko, A.~Yachmenev, W.~Thiel, J.~Tennyson, {ExoMol
  molecular line lists XXII. The rotation-vibration spectrum of silane up to
  1200\,K}, Mon. Not. R. Astron. Soc. 471 (2017) 5025--5032.
\newblock \href {https://doi.org/10.1093/mnras/stx1952}
  {\path{doi:10.1093/mnras/stx1952}}.

\bibitem{jt776}
M.~Gorman, S.~N. Yurchenko, J.~Tennyson, {ExoMol Molecular linelists -- XXXVI.
  $X$~$^{2}\Pi$ -- $X$~$^{2}\Pi$ and $A$~$^{2}\Sigma^{+}$ -- $X$~$^{2}\Pi$
  transitions of SH}, Mon. Not. R. Astron. Soc. 490 (2019) 1652--1665.
\newblock \href {https://doi.org/10.1093/mnras/stz2517/5565070}
  {\path{doi:10.1093/mnras/stz2517/5565070}}.

\bibitem{jt563}
E.~J. Barton, S.~N. Yurchenko, J.~Tennyson, {ExoMol Molecular linelists -- II.
  The ro-vibrational spectrum of SiO}, Mon. Not. R. Astron. Soc. 434 (2013)
  1469--1475.

\bibitem{jt724}
A.~Upadhyay, E.~K. Conway, J.~Tennyson, S.~N. Yurchenko, {ExoMol Molecular
  linelists -- XXV: A hot line list for silicon sulphide, SiS}, Mon. Not. R.
  Astron. Soc. 477 (2018) 1520--1527.
\newblock \href {https://doi.org/10.1093/mnras/sty998}
  {\path{doi:10.1093/mnras/sty998}}.

\bibitem{jt626}
J.~Tennyson, S.~N. Yurchenko, {The ExoMol project: Software for computing
  molecular line lists}, Intern. J. Quantum Chem. 117 (2017) 92--103.
\newblock \href {https://doi.org/10.1002/qua.25190}
  {\path{doi:10.1002/qua.25190}}.

\bibitem{00Jensen.cluster}
P.~Jensen, {An introduction to the theory of local mode vibrations}, Mol. Phys.
  98~(17) (2000) 1253--1285.
\newblock \href {https://doi.org/10.1080/002689700413532}
  {\path{doi:10.1080/002689700413532}}.

\bibitem{10HaTeKo.ai}
C.~H{\"a}tig, D.~P. Tew, A.~K{\"o}hn, {Communications: Accurate and efficient
  approximations to explicitly correlated coupled-cluster singles and doubles,
  {CCSD-F12}}, J. Chem. Phys. 132~(23) (2010) 231102.
\newblock \href {https://doi.org/10.1063/1.3442368}
  {\path{doi:10.1063/1.3442368}}.

\bibitem{08PeAdWe.ai}
K.~A. Peterson, T.~B. Adler, H.-J. Werner, {Systematically convergent basis
  sets for explicitly correlated wavefunctions: The atoms {H, He, B--Ne}, and
  {Al-Ar}}, J. Chem. Phys. 128~(8) (2008) 084102.
\newblock \href {https://doi.org/10.1063/1.2831537}
  {\path{doi:10.1063/1.2831537}}.

\bibitem{04TenNo.ai}
S.~Ten-No, Initiation of explicitly correlated {S}later-type geminal theory,
  Chem. Phys. Lett. 398~(1-3) (2004) 56--61.
\newblock \href {https://doi.org/10.1016/j.cplett.2004.09.041}
  {\path{doi:10.1016/j.cplett.2004.09.041}}.

\bibitem{09HiPeKn.ai}
J.~G. Hill, K.~A. Peterson, G.~Knizia, H.-J. Werner, {Extrapolating {MP2} and
  {CCSD} explicitly correlated correlation energies to the complete basis set
  limit with first and second row correlation consistent basis sets}, J. Chem.
  Phys. 131~(19) (2009) 194105.
\newblock \href {https://doi.org/10.1063/1.3265857}
  {\path{doi:10.1063/1.3265857}}.

\bibitem{08YoPexx.ai}
K.~E. Yousaf, K.~A. Peterson, Optimized auxiliary basis sets for explicitly
  correlated methods, J. Chem. Phys. 129~(18) (2008) 184108.
\newblock \href {https://doi.org/10.1063/1.3009271}
  {\path{doi:10.1063/1.3009271}}.

\bibitem{02Weigend.ai}
F.~Weigend, {A fully direct RI-HF algorithm: Implementation, optimised
  auxiliary basis sets, demonstration of accuracy and efficiency}, Phys. Chem.
  Chem. Phys. 4 (2002) 4285--4291.
\newblock \href {https://doi.org/10.1039/B204199P}
  {\path{doi:10.1039/B204199P}}.

\bibitem{05Hattig.ai}
C.~H{\"{a}}ttig, {Optimization of auxiliary basis sets for RI-MP2 and RI-CC2
  calculations: Core-valence and quintuple-zeta basis sets for H to Ar and
  QZVPP basis sets for Li to Kr}, Phys. Chem. Chem. Phys. 7 (2005) 59--66.
\newblock \href {https://doi.org/10.1039/b415208e}
  {\path{doi:10.1039/b415208e}}.

\bibitem{MOLPRO}
H.-J. Werner, P.~J. Knowles, G.~Knizia, F.~R. Manby, M.~Sch\"utz, Molpro: a
  general-purpose quantum chemistry program package, WIREs Comput. Mol. Sci. 2
  (2012) 242--253.
\newblock \href {https://doi.org/10.1002/wcms.82} {\path{doi:10.1002/wcms.82}}.

\bibitem{01TyTaSc.H2S}
V.~G. Tyuterev, S.~A. Tashkun, D.~W. Schwenke, An accurate isotopically
  invariant potential function of the hydrogen sulphide molecule, Chem. Phys.
  Lett. 348 (2001) 223--234.
\newblock \href {https://doi.org/10.1016/S0009-2614(01)01093-4}
  {\path{doi:10.1016/S0009-2614(01)01093-4}}.

\bibitem{97PaScxx.H2O}
H.~Partridge, D.~W. Schwenke, {The determination of an accurate isotope
  dependent potential energy surface for water from extensive ab initio
  calculations and experimental data}, J. Chem. Phys. 106 (1997) 4618--4639.
\newblock \href {https://doi.org/10.1063/1.473987}
  {\path{doi:10.1063/1.473987}}.

\bibitem{03Watson.methods}
J.~K.~G. Watson, Robust weighting in least-square fits, J. Mol. Spectrosc. 219
  (2003) 326--328.
\newblock \href {https://doi.org/10.1016/S0022-2852(03)00100-0}
  {\path{doi:10.1016/S0022-2852(03)00100-0}}.

\bibitem{88Jensen.CH2}
P.~Jensen, Calculation of rotation-vibration linestrengths for triatomic
  molecules using a variational approach: Application to the fundamental bands
  of {CH2}, J. Mol. Spectrosc. 132 (1988) 429 -- 457.
\newblock \href
  {https://doi.org/http://dx.doi.org/10.1016/0022-2852(88)90338-4}
  {\path{doi:http://dx.doi.org/10.1016/0022-2852(88)90338-4}}.

\bibitem{93JoJexx.H2O}
U.~G. J{\o}rgensen, P.~Jensen, {The dipole-moment surface and the vibrational
  transition moments of {H$_2$O}}, J. Mol. Spectrosc. 161 (1993) 219--242.
\newblock \href {https://doi.org/10.1006/jmsp.1993.1228}
  {\path{doi:10.1006/jmsp.1993.1228}}.

\bibitem{jt607}
A.~A.~A. Azzam, L.~Lodi, S.~N. Yurchenko, J.~Tennyson, {The dipole moment
  surface for hydrogen sulfide H$_{2}$S}, J. Quant. Spectrosc. Radiat. Transf.
  161 (2015) 41--49.
\newblock \href {https://doi.org/10.1016/j.jqsrt.2015.03.029}
  {\path{doi:10.1016/j.jqsrt.2015.03.029}}.

\bibitem{TROVE}
S.~N. Yurchenko, W.~Thiel, P.~Jensen, Theoretical {ROV}ibrational {E}nergies
  ({TROVE}): {A} robust numerical approach to the calculation of rovibrational
  energies for polyatomic molecules, J. Mol. Spectrosc. 245 (2007) 126--140.
\newblock \href {https://doi.org/10.1016/j.jms.2007.07.009}
  {\path{doi:10.1016/j.jms.2007.07.009}}.

\bibitem{15YaYuxx.method}
A.~Yachmenev, S.~N. Yurchenko, {Automatic differentiation method for numerical
  construction of the rotational-vibrational Hamiltonian as a power series in
  the curvilinear internal coordinates using the Eckart frame}, J. Chem. Phys.
  143 (2015) 014105.
\newblock \href {https://doi.org/10.1063/1.4923039}
  {\path{doi:10.1063/1.4923039}}.

\bibitem{17YuYaOv.methods}
S.~N. Yurchenko, A.~Yachmenev, R.~I. Ovsyannikov,
  \href{http://dx.doi.org/10.1021/acs.jctc.7b00506}{{Symmetry adapted
  ro-vibrational basis functions for variational nuclear motion: TROVE
  approach}}, J. Chem. Theory Comput. 13~(9) (2017) 4368--4381.
\newblock \href
  {http://arxiv.org/abs/http://dx.doi.org/10.1021/acs.jctc.7b00506}
  {\path{arXiv:http://dx.doi.org/10.1021/acs.jctc.7b00506}}, \href
  {https://doi.org/10.1021/acs.jctc.7b00506}
  {\path{doi:10.1021/acs.jctc.7b00506}}.
\newline\urlprefix\url{http://dx.doi.org/10.1021/acs.jctc.7b00506}

\bibitem{24Numerov.method}
B.~V. Noumerov, A method of extrapolation of perturbations, Mon. Not. R.
  Astron. Soc. 84 (1924) 592--602.
\newblock \href {https://doi.org/10.1093/mnras/84.8.592}
  {\path{doi:10.1093/mnras/84.8.592}}.

\bibitem{61Cooley.method}
J.~W. Cooley, An improved eigenvalue corrector formula for solving the
  {S}chr\"{o}dinger equation for central fields, Math. Comp. 15 (1961)
  363--374.
\newblock \href {https://doi.org/10.1090/S0025-5718-1961-0129566-X}
  {\path{doi:10.1090/S0025-5718-1961-0129566-X}}.

\bibitem{18ChJeYu.C2H2}
K.~L. Chubb, P.~Jensen, S.~N. Yurchenko, {Symmetry adaptation of the
  rotation-vibration theory for linear molecules}, Symmetry 10~(5) (2018) 137.
\newblock \href {https://doi.org/10.3390/sym10050137}
  {\path{doi:10.3390/sym10050137}}.

\bibitem{jt730}
K.~L. Chubb, A.~Yachmenev, J.~Tennyson, S.~N. Yurchenko, Treating linear
  molecule {HCCH} in calculations of rotation-vibration spectra, J. Chem. Phys.
  149 (2018) 014101.
\newblock \href {https://doi.org/10.1063/1.5031844}
  {\path{doi:10.1063/1.5031844}}.

\bibitem{98BuJexx}
P.~R. Bunker, P.~Jensen, Molecular Symmetry and Spectroscopy, 2nd Edition, NRC
  Research Press, Ottawa, 1998.

\bibitem{jt503}
S.~N. Yurchenko, R.~J. Barber, J.~Tennyson, W.~Thiel, P.~Jensen, Towards
  efficient refinement of molecular potential energy surfaces: Ammonia as a
  case study, J. Mol. Spectrosc. 268 (2011) 123--129.
\newblock \href {https://doi.org/10.1016/j.jms.2011.04.005}
  {\path{doi:10.1016/j.jms.2011.04.005}}.

\bibitem{89YaKaHi.SiH2}
C.~Yamada, H.~Kanamori, E.~Hirota, N.~Nishiwaki, N.~Itabashi, K.~Kato, T.~Goto,
  {Detection of the silylene $\nu_2$ band by infrared diode laser kinetic
  spectroscopy}, J. Chem. Phys. 91~(8) (1989) 4582--4586.
\newblock \href {https://doi.org/10.1063/1.456746}
  {\path{doi:10.1063/1.456746}}.

\bibitem{91IsKaxx.SiH2}
H.~Ishikawa, O.~Kajimoto, {Fermi resonance and vibrational analysis of SiH$_2$
  ($\tilde{A}\,^1A_1$) based on the LIF excitation-spectra of the $\tilde{A}\,
  ^1B_1$ $(060)$ $\gets$ $\tilde{A}\,^1A_1 (0v''0)$ transitions}, J. Mol.
  Spectrosc. 150~(2) (1991) 610--619.
\newblock \href {https://doi.org/10.1016/0022-2852(91)90252-6}
  {\path{doi:10.1016/0022-2852(91)90252-6}}.

\bibitem{99HiIsxx.SiH2}
E.~Hirota, H.~Ishikawa, {The vibrational spectrum and molecular constants of
  silicon dihydride SiH$_2$ in the ground electronic state}, J. Chem. Phys.
  110~(9) (1999) 4254--4257.
\newblock \href {https://doi.org/10.1063/1.478308}
  {\path{doi:10.1063/1.478308}}.

\bibitem{02IsMuMi.SiH2}
H.~Ishikawa, Y.~Muramoto, N.~Mikami, {Stimulated emission pumping spectroscopy
  of SiH$_2$: First observation of the spin-orbit interaction between the
  $\tilde{X}\,^1A_1$ A(1) and the $\tilde{a}\, ^3B_1$ states}, J. Mol.
  Spectrosc. 216~(1) (2002) 90--97.
\newblock \href {https://doi.org/10.1006/jmsp.2002.8666}
  {\path{doi:10.1006/jmsp.2002.8666}}.

\bibitem{PGOPHER}
C.~M. Western, {PGOPHER: A program for simulating rotational, vibrational and
  electronic spectra }, J. Quant. Spectrosc. Radiat. Transf. 186 (2017)
  221--242.
\newblock \href {https://doi.org/10.1016/j.jqsrt.2016.04.010}
  {\path{doi:10.1016/j.jqsrt.2016.04.010}}.

\bibitem{03YuCaJe.PH3}
S.~N. Yurchenko, M.~Carvajal, P.~Jensen, F.~Herregodts, T.~R. Huet, {Potential
  parameters of {PH}$_3$ obtained by simultaneous fitting of ab initio data and
  experimental vibrational band origins}, Chem. Phys. 290 (2003) 59--67.
\newblock \href {https://doi.org/10.1016/S0301-0104(03)00098-3}
  {\path{doi:10.1016/S0301-0104(03)00098-3}}.

\bibitem{ExoCross}
S.~N. {Yurchenko}, A.~F. {Al-Refaie}, J.~{Tennyson}, {ExoCross: a general
  program for generating spectra from molecular line lists}, Astron. Astrophys.
  614 (2018) A131.
\newblock \href {https://doi.org/10.1051/0004-6361/201732531}
  {\path{doi:10.1051/0004-6361/201732531}}.

\bibitem{87BeGrCh.SiH2}
J.~Berkowitz, J.~P. Greene, H.~Cho, B.~Ru{\v{s}}{\v{c}}i{\'{c}},
  \href{https://doi.org/10.1063/1.452213}{{Photoionization mass spectrometric
  studies of {SiH$n$} ($n=1-4$)}}, J. Chem. Phys. 86~(3) (1987) 1235--1248.
\newblock \href {https://doi.org/10.1063/1.452213}
  {\path{doi:10.1063/1.452213}}.
\newline\urlprefix\url{https://doi.org/10.1063/1.452213}

\bibitem{98EsCaxx.SiH2}
R.~Escribano, A.~Campargue, {Absorption spectroscopy of SiH$_2$ near 640 nm},
  J. Chem. Phys. 108~(15) (1998) 6249--6257.
\newblock \href {https://doi.org/10.1063/1.476062}
  {\path{doi:10.1063/1.476062}}.

\bibitem{97JeOsKo.cluster}
P.~Jensen, G.~Osmann, I.~N. Kozin, {The Formation of Four-fold Rovibrational
  Energy Clusters in H$_2$S, H$_2$Se, and H$_2$Te}, in: D.~Papousek (Ed.),
  Advanced Series in Physical Chemistry: Vibration-Rotational Spectroscopy and
  Molecular Dynamics, Vol.~9, World Scientific Publishing Company, Singapore,
  1997, pp. 298--351.

\bibitem{72DoWaxx.cluster}
A.~J. Dorney, J.~K.~G. Watson, {Forbidden Rotational Spectra of
  Polyatomic-Molecules Stark Effects and $\Delta J = 0$ Transitions of $T_d$
  Molecules}, J. Mol. Spectrosc. 42 (1972) 135--148.
\newblock \href {https://doi.org/10.1016/0022-2852(72)90150-6}
  {\path{doi:10.1016/0022-2852(72)90150-6}}.

\bibitem{84HaPaxx.cluster}
W.~G. Harter, C.~W. Patterson, {Rotational Energy Surfaces and High-$J$
  Eigenvalue Structure of Polyatomic-Molecules}, J. Chem. Phys. 80 (1984)
  4241--4261.
\newblock \href {https://doi.org/10.1063/1.447255}
  {\path{doi:10.1063/1.447255}}.

\bibitem{88Harter.cluster}
W.~G. Harter, {Computer Graphical and Semiclassical Approaches to Molecular
  Rotations and Vibrations}, Comp. Phys. Rep. 8 (1988) 319--394.
\newblock \href {https://doi.org/10.1016/0167-7977(88)90011-1}
  {\path{doi:10.1016/0167-7977(88)90011-1}}.

\bibitem{82PaAlxx}
D.~Papou{\v s}ek, M.~R. Aliev, Molecular Vibrational-Rotational Spectra,
  Elsevier, Amsterdam, 1982.

\bibitem{88ZhPaxx.cluster}
B.~I. Zhilinsky, I.~M. Pavlichenkov, {Critical Effect in Rotational Spectra of
  Water Molecule}, Opt. Spektrosk. 64 (1988) 688--690.

\bibitem{91Lehmann.cluster}
K.~K. Lehmann, {The interaction of rotation and local mode tunneling in the
  overtone spectra of symmetrical hydrides}, J. Chem. Phys. 95~(4) (1991)
  2361--2370.
\newblock \href {https://doi.org/10.1063/1.460942}
  {\path{doi:10.1063/1.460942}}.

\bibitem{93KoKlJe.cluster}
I.~N. Kozin, S.~Klee, P.~Jensen, O.~L. Polyansky, I.~M. Pavlichenkov, {The
  Far-Infrared Fourier-Transform Spectrum of H$_2$Se}, J. Mol. Spectrosc. 158
  (1993) 409--422.
\newblock \href {https://doi.org/10.1006/jmsp.1993.1085}
  {\path{doi:10.1006/jmsp.1993.1085}}.

\bibitem{94KoJexx.cluster}
I.~N. Kozin, P.~Jensen, {Fourfold Clusters of Rovibrational Energy-Levels for
  H$_2$S Studied With a Potential-Energy Surface Derived From Experiment}, J.
  Mol. Spectrosc. 163 (1994) 483--509.
\newblock \href {https://doi.org/10.1006/jmsp.1994.1041}
  {\path{doi:10.1006/jmsp.1994.1041}}.

\bibitem{96KoPaxx.cluster}
I.~N. Kozin, I.~M. Pavlichenkov, {Bifurcation in rotational spectra of
  nonlinear AB$_2$ molecules}, J. Chem. Phys. 104 (1996) 4105--4113.
\newblock \href {https://doi.org/10.1063/1.471223}
  {\path{doi:10.1063/1.471223}}.

\bibitem{96KoJePo.cluster}
I.~N. Kozin, P.~Jensen, O.~Polanz, S.~Klee, L.~Poteau, J.~Demaison, {The
  Rotational Spectrum of H$_2$Te}, J. Mol. Spectrosc. 180~(2) (1996) 402--413.

\bibitem{97GoPaJe.cluster}
P.~G\'{o}mez, L.~Pacios, P.~Jensen,
  \href{http://www.sciencedirect.com/science/article/pii/S0022285297974348}{{Fourfold
  Clusters of Rovibrational Energies in H$_2$Po Studied with an \textit{Ab
  Initio} Potential Energy Function}}, J. Mol. Spectrosc. 186~(1) (1997) 99 --
  104.
\newblock \href {https://doi.org/https://doi.org/10.1006/jmsp.1997.7434}
  {\path{doi:https://doi.org/10.1006/jmsp.1997.7434}}.
\newline\urlprefix\url{http://www.sciencedirect.com/science/article/pii/S0022285297974348}

\bibitem{93KoJexx.cluster}
I.~N. Kozin, P.~Jensen, {Fourfold Clusters of Rovibrational Energy-Levels in
  the Fundamental Vibrational-States if H$_2$Se}, J. Mol. Spectrosc. 161 (1993)
  186--207.
\newblock \href {https://doi.org/10.1006/jmsp.1993.1226}
  {\path{doi:10.1006/jmsp.1993.1226}}.

\bibitem{93JeKozz.cluster}
P.~Jensen, I.~N. Kozin, {The Potential-Energy Surface for the Electronic
  Ground-State of H$_2$Se Derived from Experiment}, J. Mol. Spectrosc. 160
  (1993) 39--57.
\newblock \href {https://doi.org/10.1006/jmsp.1993.1155}
  {\path{doi:10.1006/jmsp.1993.1155}}.

\bibitem{95FlCaBu.cluster}
J.~M. Flaud, C.~Camy-Peyret, H.~Burger, P.~Jensen, I.~N. Kozin,
  {Experimental-Evidence for the Formation of Fourfold Rovibrational Energy
  Clusters in the $v_1/v_3$ Vibrational-States of H$_2^{80}$Se}, J. Mol.
  Spectrosc. 172 (1995) 126--134.
\newblock \href {https://doi.org/10.1006/jmsp.1995.1161}
  {\path{doi:10.1006/jmsp.1995.1161}}.

\bibitem{88PaZhxx.cluster}
I.~M. Pavlichenkov, B.~I. Zhilinski\'{i}, Critical phenomena in rotational
  spectra, Ann. Phys. 184 (1988) 1--32.
\newblock \href {https://doi.org/10.1016/0003-4916(88)90268-0}
  {\path{doi:10.1016/0003-4916(88)90268-0}}.

\bibitem{05YuThPa.PH3}
S.~N. Yurchenko, W.~Thiel, S.~Patchkovskii, P.~Jensen, {Theoretical evidence
  for the formation of rotational energy level clusters in the vibrational
  ground state of PH$_3$}, Phys. Chem. Chem. Phys. 7 (2005) 573--582.
\newblock \href {https://doi.org/10.1039/b418073a}
  {\path{doi:10.1039/b418073a}}.

\bibitem{06YuThJe.SbH3}
S.~N. Yurchenko, W.~Thiel, P.~Jensen, Rotational energy cluster formation in
  {XY$_3$} molecules: Excited vibrational states of {BiH$_3$} and {SbH$_3$}, J.
  Mol. Spectrosc. 240 (2006) 174--187.
\newblock \href {https://doi.org/10.1016/j.jms.2006.10.002}
  {\path{doi:10.1016/j.jms.2006.10.002}}.

\bibitem{90Makarewicz.cluster}
J.~Makarewicz, \href{https://doi.org/10.1080/00268979000100681}{Semiclassical
  and quantum mechanical pictures of the ro-vibrational motion of triatomic
  molecules}, Mol. Phys. 69~(5) (1990) 903--921.
\newblock \href
  {http://arxiv.org/abs/https://doi.org/10.1080/00268979000100681}
  {\path{arXiv:https://doi.org/10.1080/00268979000100681}}, \href
  {https://doi.org/10.1080/00268979000100681}
  {\path{doi:10.1080/00268979000100681}}.
\newline\urlprefix\url{https://doi.org/10.1080/00268979000100681}

\bibitem{04BuJexx2.method}
P.~R. Bunker, P.~Jensen, Chirality in rotational energy level clusters, J. Mol.
  Spectrosc. 228 (2004) 640--644.
\newblock \href {https://doi.org/10.1016/j.jms.2004.02.027}
  {\path{doi:10.1016/j.jms.2004.02.027}}.

\bibitem{18OwYaKu.cluster}
A.~Owens, A.~Yachmenev, J.~K\"{u}pper,
  \href{https://doi.org/10.1021/acs.jpclett.8b01689}{Coherent control of the
  rotation axis of molecular superrotors}, J. Phys. Chem. Lett. 9~(15) (2018)
  4206--4209.
\newblock \href {https://doi.org/10.1021/acs.jpclett.8b01689}
  {\path{doi:10.1021/acs.jpclett.8b01689}}.
\newline\urlprefix\url{https://doi.org/10.1021/acs.jpclett.8b01689}

\bibitem{18OwYaYu.PH3}
A.~Owens, A.~Yachmenev, S.~N. Yurchenko, J.~K\"{u}pper, {Climbing the
  rotational ladder to chirality}, Phys. Rev. Lett. 121 (2018) 193201.
\newblock \href {https://doi.org/10.1103/PhysRevLett.121.193201}
  {\path{doi:10.1103/PhysRevLett.121.193201}}.

\end{thebibliography}

\label{lastpage}

\end{document}